\documentclass[aip]{revtex4-1}
\usepackage{amsmath}
\usepackage{amssymb}
\usepackage{color}
\usepackage{graphicx}
\usepackage{bm}
\usepackage{epstopdf}
\usepackage{ulem}
\newcommand{\Nu}{{\rm Nu}}
\graphicspath{{./figures/}}

\begin{document}


\title{Evolutionary clustering of Lagrangian trajectories in turbulent Rayleigh-B\'{e}nard convection flows} 



\author{Christiane Schneide}
\affiliation{Institute of Mathematics and its Didactics, Leuphana Universit\"at L\"uneburg, D-21335 L\"uneburg, Germany}

\author{Philipp P. Vieweg}
\affiliation{Institute of Thermodynamics and Fluid Mechanics, Technische Universit\"at Ilmenau, D-98684 Ilmenau, Germany}

\author{J\"{o}rg Schumacher}
\altaffiliation[Also at ]{Tandon School of Engineering, New York University, New York, NY 11201, USA.}
\affiliation{Institute of Thermodynamics and Fluid Mechanics, Technische Universit\"at Ilmenau, D-98684 Ilmenau, Germany}

\author{Kathrin Padberg-Gehle}
\email{kathrin.padberg-gehle@leuphana.de}
\affiliation{Institute of Mathematics and its Didactics, Leuphana Universit\"at L\"uneburg, D-21335 L\"uneburg, Germany}

\date{\today}

\begin{abstract}
We explore the transport mechanisms of heat in two- and three-dimensional turbulent convection flows by means of the long-term evolution of Lagrangian coherent sets. They are obtained from the spectral clustering of trajectories of massless fluid tracers that are advected in the flow. Coherent sets result from trajectories that stay closely together under the dynamics of the turbulent flow. For longer times, they are always destroyed by the intrinsic turbulent dispersion of material transport. Here, this constraint is overcome by the application of evolutionary clustering algorithms that add a time memory to the coherent set detection and allow individual trajectories to leak in or out of evolving clusters. Evolutionary clustering thus also opens the possibility to monitor the splits and mergers of coherent sets. These rare dynamic events leave clear footprints in the evolving eigenvalue spectrum of the Laplacian matrix of the trajectory network in both convection flows. The Lagrangian trajectories reveal the individual pathways of convective heat transfer across the fluid layer. We identify the long-term coherent sets as those fluid flow regions that contribute least to heat transfer. Thus, our evolutionary framework defines a complementary perspective on the slow dynamics of turbulent superstructure patterns in convection flows that were recently discussed in the Eulerian frame of reference. The presented framework might be well suited for studies in natural flows which are typically based on sparse information from drifters and probes.
\end{abstract}


\maketitle 

\begin{quotation} 
The transfer of heat remains one of the central questions of turbulent convection. Our present understanding of its coupling to the material transport in fully turbulent flows over longer times is still incomplete. Here, we shed new light on this point by evolutionary spectral clustering in Lagrangian networks that are formed by trajectories of fluid particles which are advected in the flow. Our framework identifies the coherent sets in two- and three-dimensional Rayleigh-B\'{e}nard convection that stay together for longer times, detects their splits and mergers, and reveals their role as those spatial regions that contribute least to the global heat transfer from the bottom to the top. This paves the way for new sparse robust investigation methods of turbulent transport in natural flows that complement the standard analyses in the Eulerian frame of reference.
\end{quotation}

\section{\label{sec:Intro}Introduction}
Turbulent thermal convection is the essential mechanism by which heat is transported across extended layers or in closed vessels \cite{Ahlers2009,Chilla2012} with numerous geophysical \cite{Stevens2005}, astrophysical \cite{Schumacher2020} and technological \cite{Kelley2018} applications. The description of convection in the Lagrangian frame of reference, which is always connected to the material transport, opens the perspective of a more precise quantification of the complex spatio-temporal pathways that heat takes through the flow and thus for a classification into spatial regions that contribute more or less to its transfer. Lagrangian transport and mixing processes have been intensively studied by means of mathematical and computational tools from dynamical systems theory \cite{Allshouse2015,Haller2015,Hadjighasem2017}. These methods have been successfully applied to the study of mass transport in several systems, such as marine ecosystems \cite{Tew2009}, cardiovascular flows \cite{Shadden2008}, astrodynamical systems \cite{Gawlik2009}, or geophysical problems including predictions of oceanic pathways of pollution patterns \cite{Olascoaga2012} and marine debris \cite{Zambianchi2017}.

At the core of Lagrangian approaches is the concept of a Lagrangian coherent set \cite{FroylandLloydSan2010,Froyland2013,Allshouse2015,Karrasch2020}, representing a regularly shaped region in the fluid volume that only weakly mixes with its surrounding. The boundaries of such regions can be identified within a geometric approach, where Lagrangian coherent structures (LCS) represent minimal curves or surfaces that enclose coherent sets \cite{Haller2015}. The LCS framework makes use of local stretching properties of the underlying flow such as measured by finite-time Lyapunov exponents (FTLE) and has been recently extended to the case of weakly diffusive transport \cite{Haller2018} and to transport of other quantities \cite{Balasuriya2018}. Coherent sets were originally introduced within a probabilistic approach based on transfer operators \cite{FroylandLloydSan2010,Froyland2013} and they have been characterized as sets which possess a minimal boundary-to-volume ratio for the dynamics by means of spectral properties of a related dynamic Laplacian operator \cite{Froyland2015}. An extension of this framework allows for studying the emergence and decay of coherent sets \cite{Froyland2021}. Recent approaches make use of spatio-temporal clustering algorithms applied to Lagrangian trajectory data \cite{FroylandPadberg2015,Hadjighasem2016,Banisch2017,Schlueter2017,Padberg2017,Schneide2018,Vieweg2021}. These unsupervised machine learning algorithms \cite{Brunton2020} aim at identifying coherent sets as groups of trajectories that remain close to each other in the time interval under investigation and can deal with sparse data and gaps in the observation. Coherent sets are also connected to the turbulent superstructures of convection \cite{Stevens2018,Pandey2018,Green2020,Krug2020,Vieweg2021a}, a gradually evolving large-scale skeleton detected in fully turbulent flows in the complementary Eulerian frame of reference \cite{Pandey2018} forming the backbone of turbulent heat transport \cite{Fonda2019}.

Conceptually, all Lagrangian approaches are tailored to identify \textit{material structures} characterized by minimal mass transport. Such a property is not expected to hold for fully turbulent flows. Turbulent dispersion dissolves larger coherent sets quickly. Coherence in this material sense rather holds for smaller-scale features such as intense vortices, which all of these approaches will preferentially identify. Moreover, the trajectory-based clustering approaches typically rely on a spatio-temporal similarity measure, which aggregates the spatio-temporal relationship between two trajectories into a single number. These similarity measures quickly become meaningless due to turbulent dispersion, causing Lagrangian approaches to fail for long-term investigations in fully turbulent flows. One first attempt to overcome this problem is to follow coherent sets from short-time computations by means of a multiple object tracking method \cite{MacMillan2020}. However, such procedure strongly relies on the capability of the underlying clustering algorithm to identify reliable structures. This sets the major motivation for the present work. 

Here, we extend the trajectory-based network approach \cite{Padberg2017} to an evolutionary spectral clustering framework \cite{Chi2007} which blends instantaneous analyses with a dynamic short-term memory. In this way, the formerly required material property of the sets of interest is relaxed thus opening doors to study the long-term evolution of Lagrangian coherent sets in two- and three-dimensional turbulent flows. The result are persistent sets with altering components, which we extract via a sparse eigenbasis approximation methodology \cite{Froyland2019} from the network Laplacian matrix. Thereby, we are able to restrict the clustering to sets which are at the center regions of the convection rolls in turbulent convection. We can then study the long-term evolution of these structures including mergers and splits - which significantly extends previous results \cite{Schneide2018,Vieweg2021} - and confirm that the coherent sets contribute least to the turbulent heat transport \cite{Vieweg2021}.
Additionally, these splits and mergers can be clearly identified by fast switches of the gaps in the spectrum of the network Laplacian. We thus present a method to study the slow dynamics of turbulent superstructures in convection from a Lagrangian view. 

The outline of the manuscript is as follows. Section II introduces the numerical simulation model of the convection flow together with important dimensionless parameters that characterize the turbulent transport. Sections III and IV discuss in brief the concept of Lagrangian coherent sets and the evolutionary clustering, respectively. Sections V and VI summarize the results for the two- and three-dimensional cases. We conclude the work in Section VII and provide further trajectory analysis details in two appendices.  

\section{Two- and three-dimensional Rayleigh-B\'{e}nard convection models}
We study Lagrangian coherent sets in thermal convection flows in a two-dimensional (2d) or three-dimensional (3d) Cartesian domain, which is uniformly heated from below and cooled from above. The horizontal extension $L$ of the domains is in both cases significantly larger than their vertical extension $H$, which is quantified by the aspect ratio $\Gamma = L / H$. The two-dimensional (2d) flow case is studied in a closed domain $x\in [-4,4]$ and $z\in [0,1]$. In the three-dimensional (3d) case, we consider $x,y \in [-8, 8]$ and the vertical coordinate $z$ in the same range as in the 2d case. All coordinates are given in units of the layer height $H$.

\begin{figure}[htb]
    \begin{center}
    \includegraphics[width=0.65\textwidth]{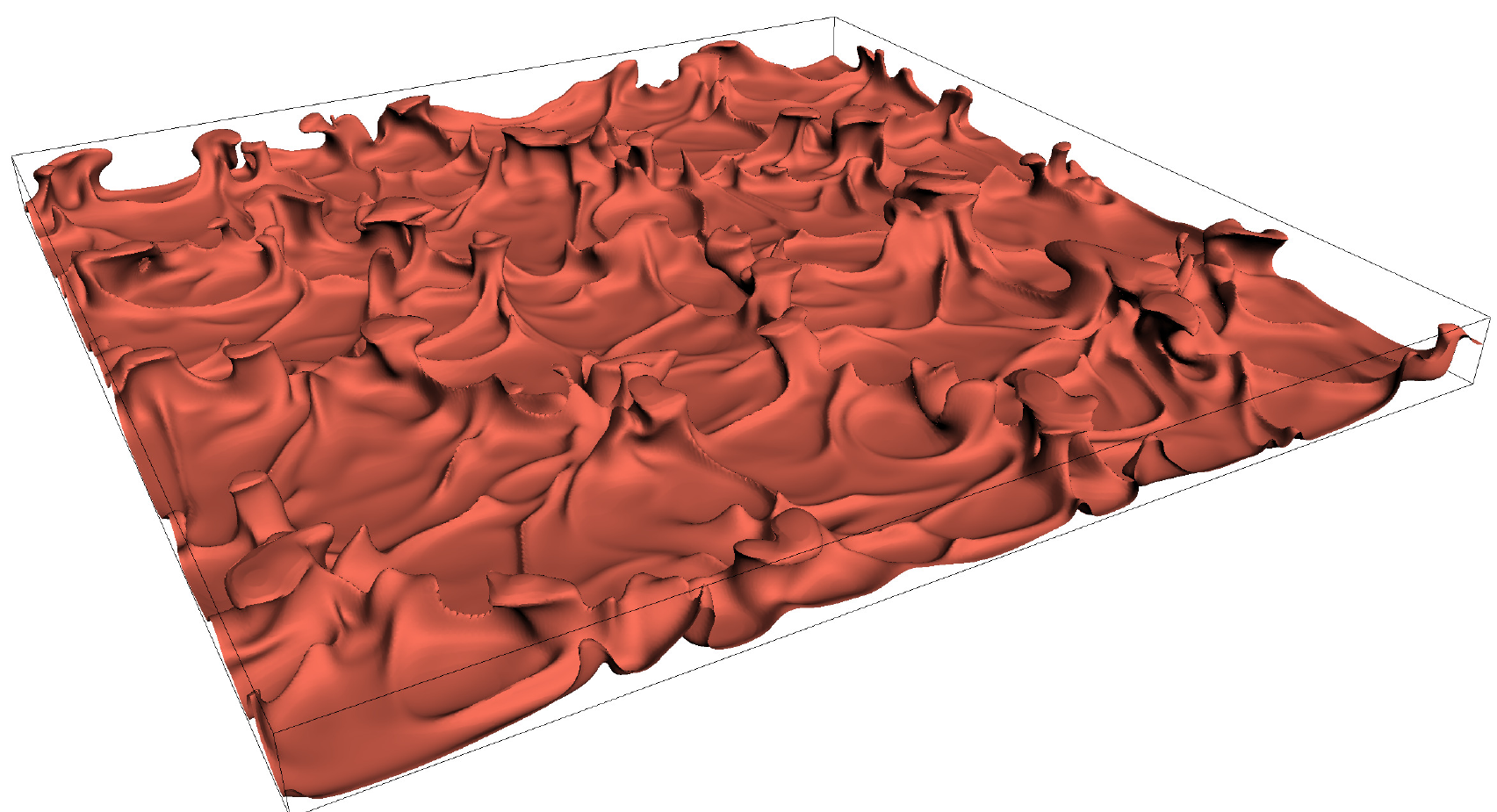}
    \caption{Perspective view of the instantaneous iso-surface at $T(x,y,z)=0.7$ in turbulent Rayleigh-B\'enard convection. The system parameters $\text{Pr}=0.7$, $\text{Ra}=10^{5}$ and $\Gamma=16$. This iso-surface shows regions of up-welling fluid flow, indicating regions of large convective heat flux. (Leaking) Lagrangian coherent sets can be found between this hot iso-surface and its cold counterpart (down-welling thermal plumes, not shown), thus reducing the overall heat transport.}
    \label{fig:isosurface}
    \end{center}
\end{figure}

The underlying equations of motion in the Boussinesq approximation are solved in both cases by a direct numerical simulation applying the spectral element method Nek5000 \cite{Scheel2013}. The equations are made dimensionless by the layer height $H$, wall-to-wall temperature difference $\Delta T$, and free-fall velocity $U_{f} = \sqrt{g \alpha \Delta T H}$ with the acceleration due to gravity $g$ and the isobaric expansion coefficient $\alpha$. Consequently times are measured by the free-fall time $T_f$ which is given by
\begin{equation}
    T_f=\frac{H}{U_f}=\sqrt{\frac{H}{g\alpha\Delta T}}\,.
\end{equation}
The equations of motion in dimensionless form follow to
\begin{align}
\label{eq:continuity_equation}
\nabla \cdot {\bm u} &= 0\,,\\
\label{eq:Navier-Stokes_equation}
\frac{\partial {\bm u}}{\partial t} + \left( {\bm u} \cdot \nabla \right) {\bm u} &= - \nabla p + \sqrt{\frac{\text{Pr}}{\text{Ra}}} \nabla^{2} {\bm u} + T {\bm e_{z}} \,,\\
\label{eq:energy_equation}
\frac{\partial T}{\partial t} + \left( {\bm u} \cdot \nabla \right) T &= \frac{1}{\sqrt{\text{Ra Pr}}} \nabla^{2} T\,.
\end{align}
The variables ${\bm u}({\bm x},t), T({\bm x},t)$ and $p({\bm x},t)$ represent the non-dimensional velocity, temperature, and pressure fields, respectively, with ${\bm x}=(x,y,z)^T$. The Rayleigh number ${\rm Ra}$, which is given by 

\begin{equation}
    {\rm Ra} = \frac{\alpha g \Delta T H^{3}}{\nu \kappa}\,,
\end{equation}

describes the strength of thermal driving. The Prandtl number ${\rm Pr}$, which is given by
\begin{equation}
    {\rm Pr} = \frac{\nu}{\kappa}\,,
\end{equation}
measures the ratio of kinematic viscosity $\nu$ to temperature diffusivity $\kappa$, thus being a working fluid parameter. Table \ref{tab:S1_numerical_models} summarizes all important parameters and quantities for both, the two-dimensional and three-dimensional simulation. Both simulations are for the turbulent flow case. Figure \ref{fig:isosurface} shows an iso-surface snapshot of the 3d turbulent temperature field at the isolevel $T=0.7$ which illustrates clearly the rising thermal plume structures. While the spectral element code works in the Eulerian frame of reference, the advection of $N$ massless tracer particles is an analysis in the Lagrangian frame of reference. It is conducted together with the time advancement of the Boussinesq equations. The trajectories of the $N$ tracers follow 
\begin{equation}
    \dot{\bm X}_i={\bm u}({\bm X}_i,t) \quad\mbox{for}\quad i=1,\dots,N\,.
    \label{Lag}
\end{equation}
The velocity field components in (\ref{Lag}) are obtained by spectral interpolation to the tracer position $\bm X(t)$. The spectral resolution of the simulations is governed by the number of spectral elements $N_{e}$ together with the polynomial order $N_{p}$ on each spectral element and in each spatial dimension. The temporal advancement of the simulations exploits a second order backwards difference formula.

The initial conditions (ICs) of the simulations are given by a linear vertical temperature profile at equilibrium subject to a small random perturbation $\Upsilon$ of maximum order $\mathcal{O}(10^{-3})$ together with a fluid at rest, i.e., $T ({\bm x}, t = 0) = 1 - z + \Upsilon$ and ${\bm u} ({\bm x}, t = 0) = 0$. The initial transient evolution of the dynamical system towards a statistically stationary state is ignored for the fields; the analysis starts once the tracer particles are homogeneously distributed across the simulation domain. The boundary conditions (BCs) are as follows. The flows are driven by uniform temperatures (Dirichlet-type BCs) at the bottom and top plate, i.e. $T(x, y, z = 0, t) = 1$ and $T(x, y, z = 1, t) = 0$, respectively, whereas the lateral walls are thermally insulating (Neumann-type BC), i.e. $\partial T / \partial {\bm n} = 0$ where ${\bm n}$ is the normal vector of the corresponding lateral wall. The simulation domains are closed in all directions and exhibit -- just as in a real laboratory experiment -- no-slip BCs, i.e. ${\bm u} ({\bm x} = {\bm x}_{b}, t) = 0$ with domain boundaries located at ${\bm x}_{b}$. The aspect ratios are chosen large enough such that finite-domain-size effects are negligible.

The Nusselt number (in the Eulerian frame of reference) ${\rm Nu}^{E}$ is given by
\begin{equation}
    {\rm Nu}^{E} = - \Big\langle \left. \frac{\partial T}{\partial z} \right|_{z = 0} \Big\rangle_{\partial\Omega, t} = - \Big\langle \left. \frac{\partial T}{\partial z} \right|_{z = 1} \Big\rangle_{\partial\Omega, t}\,.
\end{equation}
It is the global measure of mean turbulent heat transfer from the bottom to the top with ${\rm Nu}^{E}\ge 1$. Here, $\langle \cdot \rangle_{\partial\Omega, t}$ is either a combined average with respect to time $t$ and a line at constant $z$ in 2d or time and a horizontal cross-section area $A$ in 3d. The Reynolds number ${\rm Re}$, which is given by
\begin{equation}
    {\rm Re} = \sqrt{\frac{{\rm Ra}}{{\rm Pr}} \thickspace \langle {\bm u}^{2} \rangle_{\Omega, t}}\,,
\end{equation}
is the global measure of turbulent momentum transport. Here, $\langle \cdot \rangle_{\Omega, t}$ is either a combined average with respect to time $t$ and simulation plane in 2d or time and simulation volume in 3d. In addition to the global measure of turbulent heat transport (see above), we compute for each individual tracer particle $i$ a \textit{local} Nusselt number ${\rm Nu}_{\rm local}$ in the Lagrangian frame of reference along its trajectory ${\bm X}_{i}(t)$. This Nusselt number is given by \cite{Vieweg2021}
\begin{equation}\label{eq:localNu}
    {\rm Nu}_{\rm local} \left[ {\bm X}_{i}(t) \right] = \left. \sqrt{{\rm Ra} {\rm Pr}} \thickspace u_{z} T \right|_{{\bm X}_{i}(t)} - \left. \frac{\partial T}{\partial z} \right|_{{\bm X}_{i}(t)} \,.
\end{equation}
We verified that the Lagrangian ensemble average of the time-averaged local Nusselt numbers coincides with the mean value ${\rm Nu}^{E}$ in the Eulerian frame of reference. This is documented in Table \ref{tab:S1_numerical_models}. Here, the time average is taken over the whole time interval $\mathbb{T}$ and the Lagrangian ensemble average over all tracers ($L$). This results to
\begin{equation}
\langle {\rm Nu}_{\rm local} \rangle_{\mathbb{T},L}= \frac{1}{N}\sum_{i=1}^N \langle {\rm Nu}_{\rm local} \left[ {\bm X}_{i}(t) \right] \rangle_{\mathbb{T}} \approx {\rm Nu}^{E}\,.
\end{equation}

From the table we note that the standard deviations of the Eulerian and Lagrangian Nusselt numbers differ considerably. This is caused by the ways they are computed. The comparatively large standard deviation in the Lagrangian frame represents different contributions of individual trajectories;  the smaller Eulerian value represents temporal changes in the line or area average $\langle \cdot \rangle_{\partial\Omega}$.

\begin{table}[!htb]
    \centering
    \begin{tabular}{c c c c r c r c c c c}
    \hline
    $d$   & $\Gamma$  & ${\rm Ra}$    & ${\rm Pr}$    & $N_{e}$    & $N_{p}$     & $\Delta t^{L}$  & $N$          & ${\rm Nu}^{E}$   & $\langle {\rm Nu}_{\rm local} \rangle_{\mathbb{T},L}$  & ${\rm Re}$ \\
    \hline
    $2$   & $8$       & $10^{8}$      & $7.0$         & $384$      & $11$        & $1500$   & $16,384$     & $26.4 \pm 0.9$    & $26.2\pm 9.6$    & $567.3 \pm 10.7$\\
    $3$   & $16$      & $10^{5}$      & $0.7$         & $441,000$  & $5$         & $158$    & $65,536$    & $4.1 \pm 0.1$   & $4.3\pm 1.0$         & $91.1 \pm 0.5$ \\
    \hline
    \end{tabular}
    \caption{Parameters and statistical measures of the numerical models. The dimension $d$ of the spatial domain, the aspect ratio $\Gamma$, the Rayleigh number ${\rm Ra}$, the Prandtl number ${\rm Pr}$, the number of spectral elements $N_{e}$, the polynomial order $N_{p}$ on each spectral element, the total time of Lagrangian analysis $\Delta t^{L}$, the number $N$ of massless Lagrangian tracer particles, the global Nusselt number ${\rm Nu}^{E}$ in the Eulerian frame of reference, its Lagrangian trajectory-based counterpart $\langle {\rm Nu}_{\rm local} \rangle_{\mathbb{T},L}$, as well as the global Reynolds number ${\rm Re}$ are listed. The latter three quantities are given with respect to their mean value and the corresponding standard deviation.}
    \label{tab:S1_numerical_models}
\end{table}

\begin{figure}[htb]
\begin{center}
\includegraphics[width=0.65\textwidth]{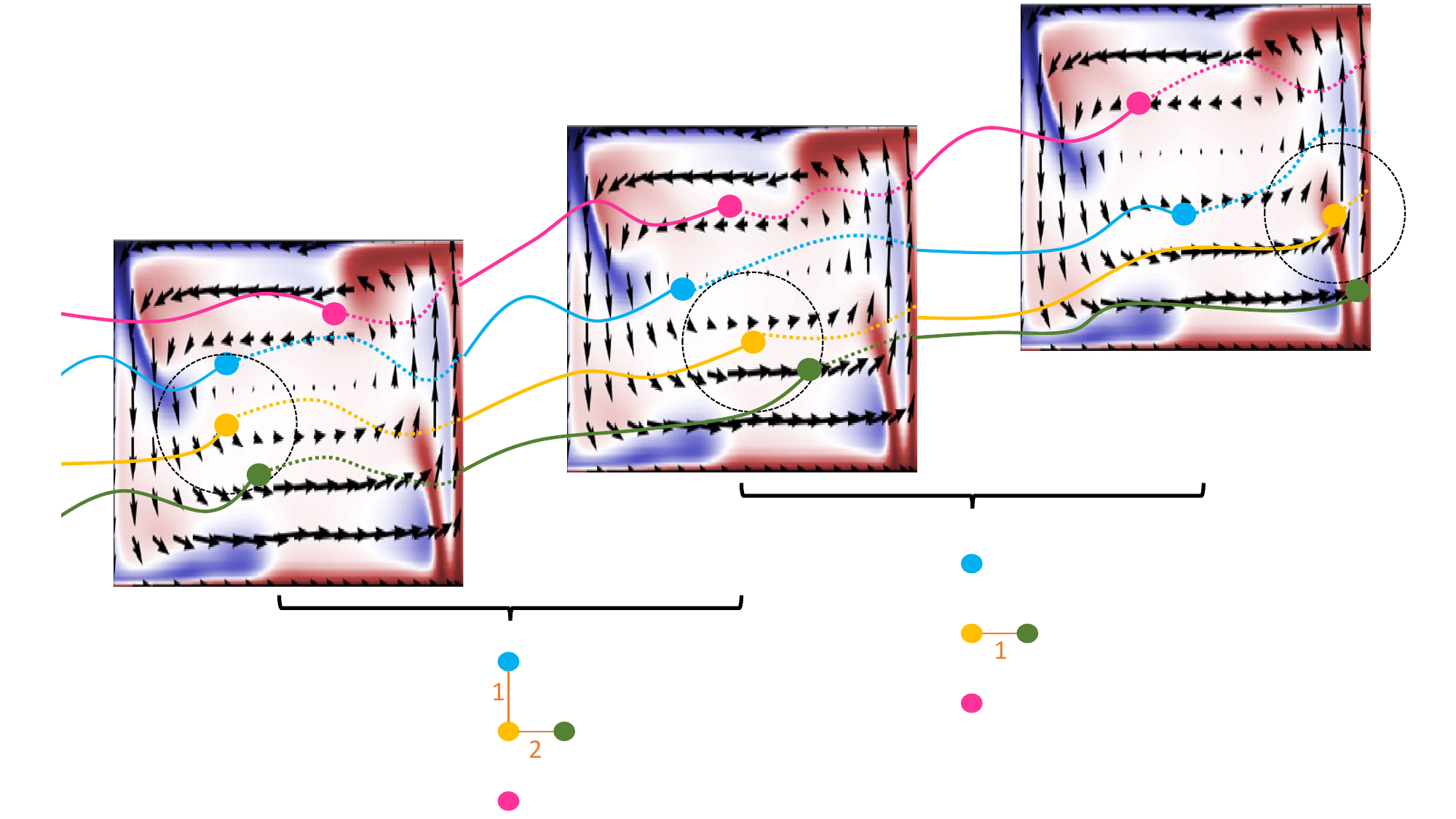}
\caption{\label{fig:scheme} Sketch of Lagrangian tracers moving in a convection cell. The black dashed ellipse represents the $\epsilon$-neighborhood of the yellow particle. Consecutive weighted networks computed over the first and second snapshot (second and third, respectively) are shown below.}
\end{center}
\end{figure}

\section{Trajectory-based Lagrangian coherent sets}
Recently, complex network approaches have been successfully applied to the study of turbulent flows \cite{Iacobello2021}.
In this work, we extend the trajectory-based network framework for the identification of coherent sets proposed in Ref.~\onlinecite{Padberg2017}, where an unweighted network was considered.  Given $N$ tracer trajectories ${\bm X}_i(t)$ at discrete time instances $t\in \mathbb{T}=\{0,1, \ldots, \Theta \}$ and a threshold $\epsilon>0$, we set up instantaneous adjacency matrices $A_t \in \{0, 1\}^N$  where $A_{ij,t}=1 $  if $\|{\bm X}_i(t)-{\bm X}_j(t)\|<\epsilon$ for $i\neq j$ and  $A_{ij,t}= 0$ otherwise. From this we obtain a network weight matrix $W=\sum_{t\in \mathbb{T}} A_t$ with components $W_{ij}$ that encode the number of $\epsilon$-close encounters of tracers ${\bm X}_i$ and ${\bm X}_j$ over the entire time span. Hence, $W_{ij}$ is large when the trajectories ${\bm X}_i$ and ${\bm X}_j$ are close. Classical network measures can be employed, such as the weighted and unweighted node degree which are given by $d_{w,i}=\sum_{j=1}^N W_{ij}$ and $d_{u,i}=\sum_{j=1}^N \overline{W}_{ij}$, respectively. Here, $\overline{W} \in \{0, 1\}^N$ is the adjacency matrix of the unweighted network. The latter measures how many different trajectories interact with ${\bm X}_i$ and can be linked to FTLE \cite{Banisch2019}. In our context, the normalized node degree, which is given by $d_i=d_{u,i}/d_{w,i}$, measures the number of different links of node ${\bm X}_i$ in relation to the total links weights. It is expected to be large in highly mixing regions, where each particle interacts with many different other ones. 
In order to analyze regions with different dynamical behavior, we seek to create a hard clustering of the trajectories. To this end, for example, one could define a threshold on the normalized node degree. Another approach is the application of spectral graph partitioning techniques, i.e.\ the partitioning of a data set that is represented by a graph based on spectral properties of a corresponding similarity matrix. Here, we use the weight matrix $W$ as similarity matrix and optimize the normalized cut (NCut), originally proposed in Ref.~\onlinecite{ShiMalik2000}, to determine the hard clustering. The cost function of the k-way normalized cut  \cite{Bach2006}, i.e., a partitioning into $k$ sets, to be minimized is given by  
\begin{equation}
{\rm Cost}_{\rm NCut} = k - {\rm Tr} \left[ X^T \left(D^{-1/2}WD^{-1/2}\right) X\right]\,,
\end{equation}
where $D \in \mathbb{R}^{N,N}$ denotes the degree matrix with $D_{ii}=d_{w,i}$. Therefore, the entry $i,j$ of the matrix $D^{-1/2}WD^{-1/2}$ is given by $W_{ij}/\sqrt{d_{w,i}d_{w,j}}$. The normalized weight matrix resembles a random walk matrix and has eigenvalues $\lambda_i \leq 1$. The cost function ${\rm Cost}_{\rm NCut}$ is minimized by the matrix $X \in \mathbb{R}^{N,k}$ if the columns of $X$ are composed of the eigenvectors associated to the $k$ largest eigenvalues of the matrix $ D^{-1/2}WD^{-1/2}$, where $k$ can be determined using a spectral gap criterion. The clustering is subsequently produced in a post-processing step using the eigenvectors of the matrix $X$. To this end, we use the sparse eigenbasis approximation (SEBA)\cite{Froyland2019}. From this, we derive a cluster membership indicator $S_{\rm max}$, which highlights particles in coherent regions (see Appendix \ref{sec:AppSEBA} for mathematical details). 
\begin{figure*}[htb]
\begin{center}
\includegraphics[width=16.5cm]{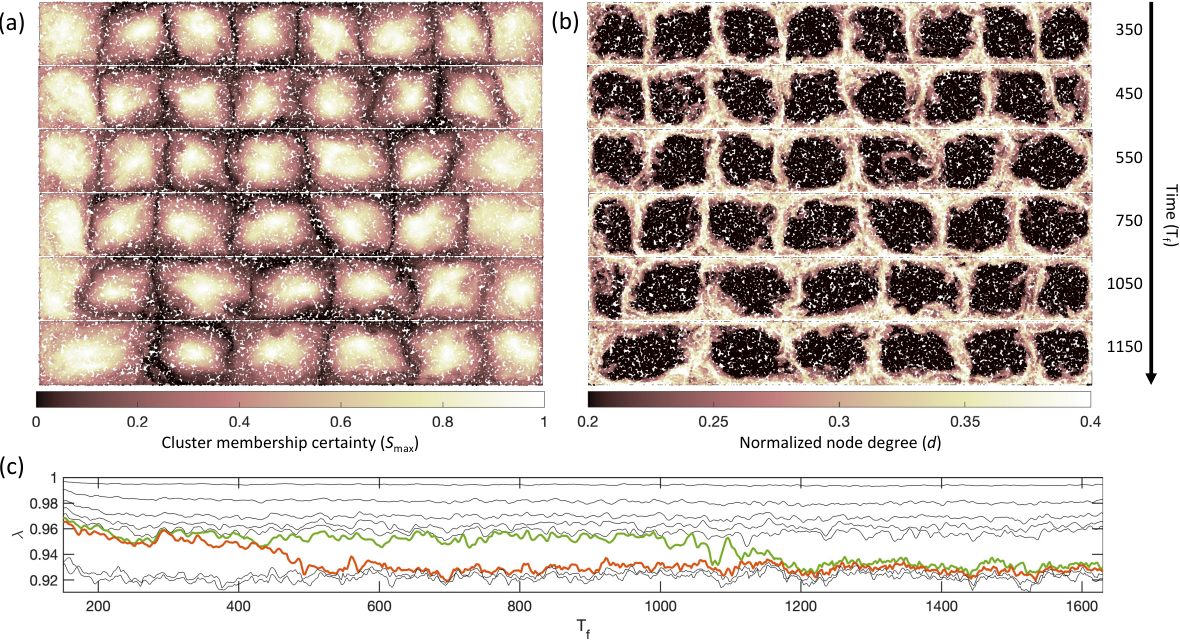}
\caption{\label{fig:evolution2d} Long-term evolution of coherent sets using the network evolution approach displays two merging events. (a) Postprocessing of the spectral information outputs a likelihood $S_{\rm max}({\bm X}_i(t))$ for each particle trajectory ${\bm X}_i$ to belong to a cluster over the time span $t\pm \frac{\tau}{2}$ (multimedia view), see Appendix \ref{sec:AppSEBA}. (b) The normalized node degree $d({\bm X}_i(t))$ gives a similar distinction of clusters and mixing regions (multimedia view). Similarly to panel (a), the whole simulation domain is visualized at 6 time instances. (c) The 10 leading eigenvalues of the matrix $D_t^{-1/2}W_t D_t^{-1/2}$ plotted versus time highlight the existence of eight coherent sets at the beginning which merge into seven and then six at about $t=480 T_f$ and $1150 T_f$, respectively; see Appendix \ref{sec:AppSpectrum} for more details on the evolution of the spectrum.}
\end{center}
\end{figure*}

\section{Evolutionary spectral clustering}
Building on the trajectory-based network representation we propose two evolutionary frameworks that allow us to study the evolution of large-scale coherent sets in turbulent flows in the Lagrangian frame of reference over long time spans.

\subsection{Evolving network approach}\label{sec:ena}
First, we consider time-dependent weight matrices $W_{t}$ that are computed over a shorter time period $\tau \ll \Theta$, centered at $t$, i.e., $\mathbb{T}_{t,\tau}=\{t-\frac{\tau}{2}, \ldots,  t+\frac{\tau}{2}\}$, via $W_{t}=\sum_{s \in \mathbb{T}_{t,\tau}} A_s$. Obviously, the weight matrix $W_{t}$ is constructed from $W_{t-1}$ by adding future information $A_{t+\frac{\tau}{2}}$ and dropping the most historical information $A_{t-1-\frac{\tau}{2}}$, see Figure~\ref{fig:scheme} for an illustration. When $\tau$ is sufficiently large, $W_t$ and thus also the matrix of the leading $k$ eigenvectors $X_t$ from the relaxed normalized cut problem can be expected to vary ``smoothly'' in time. 
  
This construction is an adoption of the sliding window approaches in the transfer operator framework \cite{Blachut2020,Ndour2021} to the trajectory-based context and thus considered as a variation of the evolutionary spectral clustering approach \textit{preserving cluster quality} (PCQ) \cite{Chi2007}. Notably, by this evolutionary construction the material property of coherent sets over the whole time span $\mathbb{T}$ is relaxed and holds strictly only over smaller time spans of length $\tau$ used for setting up $W_t$. Significant changes in the dynamics such as mergers or splits of coherent sets will leave their footprints in the time evolution of the spectral properties of $D_t^{-1/2}W_tD_t^{-1/2}$, which can be highlighted by means of the cluster membership certainty $S_{\rm max}$ or local network measures such as the normalized node degree $d$.  By $S_{\rm max}({\bm X}_i(t))$ as well as $d({\bm X}_i(t))$ we denote the respective time-dependent quantities for trajectories ${\bm X}_i$ over the time span $t\pm\frac{\tau}{2}$ centered at time $t$. 
 
\subsection{Evolving cluster approach} \label{sec:eca}
The evolving network approach works well when sufficiently strong spectral gaps exist such that short-term variations due to systematic errors or noise do not cause the recent clustering to deviate strongly from the preceding one, such as also required in Ref.~\onlinecite{MacMillan2020}. In case of a simultaneous monitoring of the evolution of several coherent sets or of sub-dominant coherent sets, one can consider a second evolutionary framework, the \textit{preserving cluster membership} (PCM) \cite{Chi2007} , which directly compares the recent clustering to the preceding clustering. This concept translates into a normalized cut problem of the form \cite{Chi2007}
\begin{equation}
{\rm Cost}_{\rm NCut,t} = k - {\rm Tr} \left[ X_t^T \widehat{W}_t X_t\right]\, ,
\end{equation}
which is solved by $X_t \in \mathbb{R}^{N,k}$ if the columns of $X_t$ are composed of the eigenvectors associated to the $k$ leading largest eigenvalues of the matrix 
\begin{equation}
    \widehat{W}_t=\alpha D_t^{-1/2}W_t D_t^{-1/2} + (1-\alpha) X_{t-1}X_{t-1}^T \,,
\end{equation}
where $X_{t-1}$ is obtained in the previous time step.  The parameter $\alpha$ ($0 \leq \alpha \leq 1$) weights the significance of the current clustering in comparison to the preceding one, the latter of which is encoded in $X_{t-1}$. 
Thus, the matrix $\widehat{W}_t$ is composed of $X_{t-1}X_{t-1}^T$ representing the previous clustering, weighted by $(1-\alpha)$, and of $D_t^{-1/2}W_t D_t^{-1/2}$ representing the current connectivity of the nodes, weighted by $\alpha$.
Note that while $W_t$ is sparse, $\widehat{W}_t$ no longer is due to $X_{t-1}X_{t-1}^T$ being a full matrix. This problem can be overcome by using a sparse approximation of the eigenspace spanned by $X_{t-1}$ from the SEBA postprocessing \cite{Froyland2019} (see Appendix \ref{sec:AppSEBA}). Thereby, the computational cost of the analysis can be reduced extremely, making the study of large trajectory networks possible. 

\begin{figure*}[htb]
\begin{center}
\includegraphics[width=16.5cm]{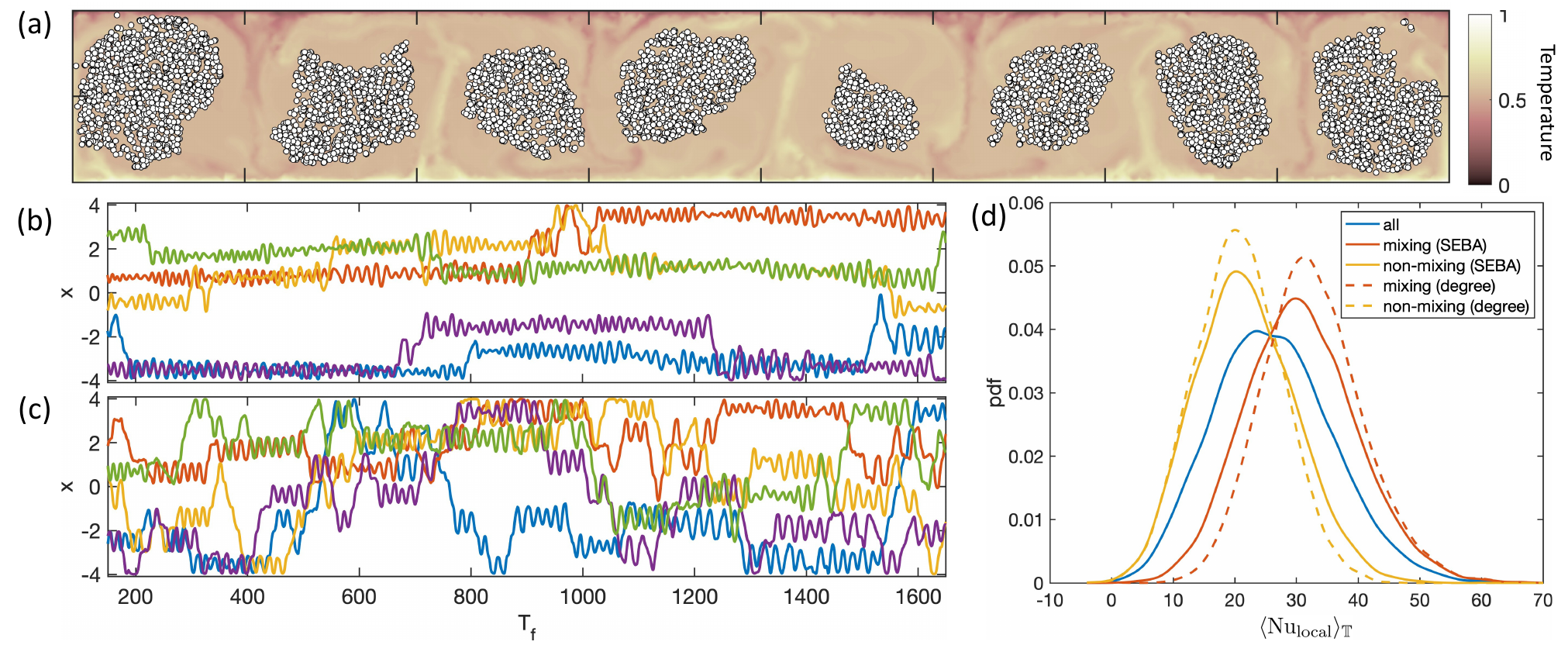}
\caption{\label{fig:results2d} Trajectory analysis in 2d case. (a) Temperature field at $t=250 T_f$ shown together with tracers with $S_{\rm max}({\bm X}_i(t)) \geq 0.5$ for $t=250 T_f$. (b) Horizontal $x$-coordinates versus time of example trajectories that mostly belong to coherent sets. (c) Horizontal $x$-coordinates of trajectories that are mostly outside the coherent sets. (d) Probability density functions (pdfs) of the time-averaged local Nusselt number $\langle \Nu_{\rm local}\rangle_{\mathbb{T}}$ averaged over $1500 T_f$. The analysis is either conducted over all trajectories or two subgroups, either with respect to the cluster membership indicator $S_{\rm max}$ obtained via SEBA \cite{Froyland2019} or to the normalized node degree $d$, see legend. All pdfs are smoothed by a Gaussian kernel.}
\end{center}
\end{figure*}

\section{Two-dimensional Rayleigh-B\'{e}nard convection flow}\label{sec:2d}
We start with the 2d RBC case where 16,384 particle trajectories are analysed after an initial phase of $150 T_f$ over a long time period of $1500 T_f$, i.e., $\mathbb{T}=\{150, 151, \ldots, 1650 \}$, where time is given in free-fall times $T_f$. With a threshold of $\epsilon = 0.1$, we set up the weighted adjacency matrices $W_t$ for the observation time $\tau=20$. Using the evolving network approach from section \ref{sec:ena}, we consider the spectral properties of $D_t^{-1/2}W_t D_t^{-1/2}$ for $t=160$ to $1640$ (in steps of $2$). Figure \ref{fig:evolution2d}(a) shows the time evolution of trajectory clusters as a result of the SEBA analysis of the leading $k$ eigenvectors, where $k\in \{6,7,8\}$ is chosen according to a spectral gap that shifts with progressing time (cf. Figure \ref{fig:evolution2d}(c)). Light regions correspond to coherent features, whereas dark regions indicate the incoherent background possessing a low $S_{\rm max}({\bm X}_i(t))$. These are the regions between the convection rolls where hot fluid rises and cold fluid descents. Starting with eight clusters, the merger of the two right-most clusters between $450$ and $550 T_f$ (more precisely, at about $t=480 T_f$, see Figure \ref{fig:evolution2d}(a) (multimedia view)), as well as a further merger of the two left-most clusters between $1050$ and $1150 T_f$, results in six final clusters that remain stable until the end of the simulation. Figure \ref{fig:evolution2d}(b) visualizes the normalized node degree $d({\bm X}_i(t))$ for the same times as panel (a), offering a similar detection of the mergers of two coherent sets (see also Figure \ref{fig:evolution2d}(b) (multimedia view)). The node degree gives -- as expected due to its relation to FTLE \cite{Banisch2019} -- a complementary picture with high values in the mixing and lower ones in the coherent regions. Figure \ref{fig:evolution2d}(c) shows the temporal evolution of the leading eigenvalues. There is a prominent spectral gap between the 8th and 9th eigenvalue until about $t=450 T_f$, where the 8th eigenvalue (red curve) becomes significantly smaller and a new gap appears between the 7th and 8th eigenvalue. The former (green curve) also drops at about $t=1150 T_f$. Thus, the mergers of the coherent sets leave their clear footprints in the spectrum, confirming previous observations using transfer operator methods \cite{Blachut2020,Ndour2021}. 

Figure \ref{fig:results2d}(a) shows the temperature field at time $t=250 T_f$. Superimposed are Lagrangian particles with a cluster membership certainty $S_{\rm max}({\bm X}_i(t))$ of at least 50\% at this time. One observes that the extracted clusters lie in between the regions of ascending hot (light colors) and descending cold (dark colors) fluid and thus form the centers of convection rolls. In Fig. \ref{fig:results2d}(b) we plot the horizontal $x$-coordinate of five selected Lagrangian tracers over time. These tracers exhibit a large average coherent set membership likelihood over the whole simulation time of $1500 T_f$ and tend to stay in the centers of convection rolls for long periods before switching to the center of a different roll. This is in contrast to particles with a cluster membership likelihood below average. These tend to switch frequently between different roles as illustrated in Fig. \ref{fig:results2d}(c).
 
To study the turbulent heat transport, we compute the time-averaged local Nusselt number $\langle \Nu_{\rm local}\rangle_{\mathbb{T}}$ for each particle trajectory (see Eq.~\ref{eq:localNu}). The additional average over all trajectories (L) results to $\langle \Nu_{\rm local}\rangle_{\mathbb{T},L}=26.2 \pm 9.6$ which is very close to the Nusselt number ${\rm Nu}^{E}$ obtained in the Eulerian frame of reference (see Table \ref{tab:S1_numerical_models}). The probability density function (pdf) of the time-averaged local Nusselt numbers is shown in Fig. \ref{fig:results2d}(d) (solid blue curve). To investigate the different contributions of coherent and complementary sets better, we compute both the time-averaged values $\langle S_{\rm max}({\bm X}_i(t))\rangle_{\mathbb{T}}$ and $\langle d({\bm X}_i(t))\rangle_{\mathbb{T}}$ for every particle trajectory as well as the respective additional averages over all trajectories $\langle S_{\rm max}\rangle_{\mathbb{T},L}=0.39$ and $\langle d\rangle_{\mathbb{T},L}=0.23$. Trajectories in coherent sets with $\langle S_{\rm max}({\bm X}_i(t))\rangle_{\mathbb{T}} \ge \langle S_{\rm max}\rangle_{\mathbb{T},L}$ exhibit a time-averaged local Nusselt number $\langle \Nu_{\rm local, c}\rangle_{\mathbb{T}} = 21.7 \pm 8.0$ (solid yellow curve, 8179 tracers), whereas the remaining trajectories contribute significantly more to the turbulent heat transport and mixing with $\langle \Nu_{\rm local,  m}\rangle_{\mathbb{T}} = 30.7 \pm 9.0$ (solid red curve, 8205 tracers). The corresponding distinction by means of the normalized node degree $\langle d({\bm X}_i(t))\rangle_{\mathbb{T}}$ results in an even stronger separation of the heat transport contributions with $20.6\pm 7.0$ (dashed yellow curve, 8699 tracers) and $32.6 \pm 8.0$ (dashed red curve, 7685 tracers), respectively. These results clearly indicate that coherent sets contribute significantly less to the turbulent heat transport with tracers trapped inside the convection rolls (cf. Figure \ref{fig:results2d}(a)) for longer times, underlining the findings of  Ref.~\onlinecite{Vieweg2021} by the present framework.
\begin{figure*}
\begin{center}
\includegraphics[width=17cm]{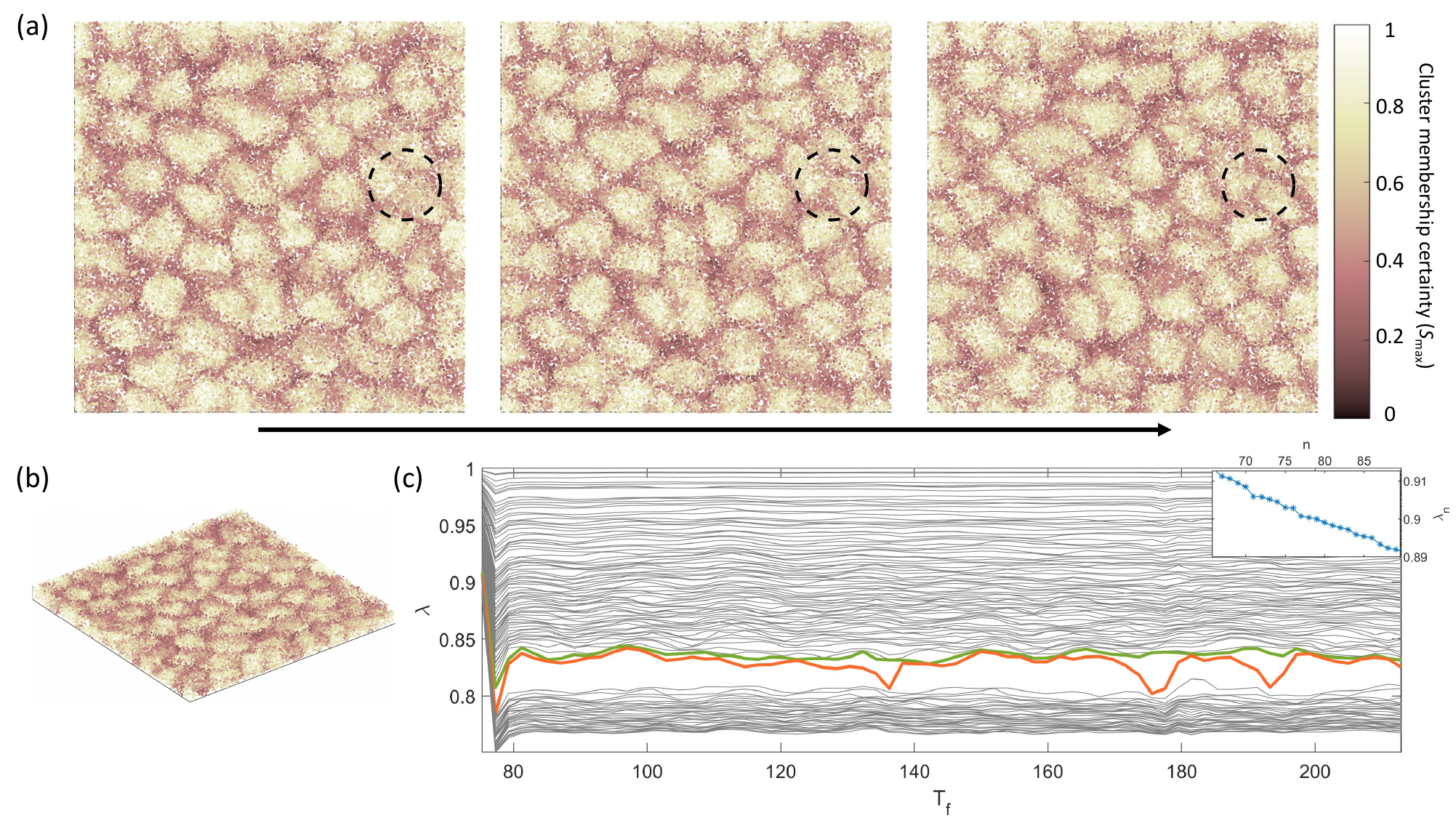}
\caption{\label{fig:evolution3d} Trajectory analysis in 3d case. (a) Top view onto the layer at $t= 128, 130$ and $132 T_f$ indicating cluster membership certainty by color. Dashed circles mark a local splitting event. (b) 3d view of first figure of panel (a). (c) Leading eigenvalues plotted versus time. Marked in green and orange are the 70th and 71st eigenvalues, respectively. Inset: 65th to 90th eigenvalues are displayed for the first evolutionary step of $\tau/10$.}
\end{center}
\end{figure*}

\section{Three-dimensional Rayleigh-B\'{e}nard convection flow}\label{sec:3d}
The analysis of the 3d data turns out to be more complex, it is based on 65,536 trajectories over a time period $158 T_f$, precisely $\mathbb{T}=\{75.3, 77.2, \ldots, 212.9 \}$. 
Here, we create a weighted network with parameters similar to the 2d case, $\tau = 19.65 T_f$ and $\epsilon \approx 0.14$. This results again in sparse adjacency matrices with edge densities\cite{Donner2010} of $\rho(\epsilon) < 10^{-4}$. The number of Lagrangian coherent sets is estimated based on physical assumptions\cite{Vieweg2021} to be approximately 80. In the first time step, centered at $t = 75.3 T_f$, we detect a large difference between the 70th and 71st eigenvalue (cf. inset Figure \ref{fig:evolution3d}(c)); therefore we take $k = 70$. In the subsequent analysis, this value will vary within $[0.9k_{t-1}, 1.1k_{t-1}]$. The large number of coherent sets in the 3d data set suggests the usage of the evolving cluster approach from section \ref{sec:eca} to stabilize the investigated coherent sets. For our study, we advance in time in steps of $\tau/10$ and take $\alpha = 0.9$. As in the two-dimensional case, we investigate the temporal evolution of the leading eigenvalues, see Fig.~\ref{fig:evolution3d}(c). The variation of the 71st eigenvalue corresponds to splits and mergers of coherent sets which are clearly highlighted in plots of $S_{\rm max}({\bm X}_i(t))$ in Fig.~\ref{fig:evolution3d}(a) at three consecutive times. A 3d image of $S_{\rm max}({\bm X}_i(t))$ for the first time instance is given in Fig.~\ref{fig:evolution3d}(b). Multiple splits and mergers of coherent sets proceed at other instances as well. If the total number of coherent sets is unchanged these events are not traceable in the spectrum.

\begin{figure}[htb]
\centering
\includegraphics[width=0.36\textwidth]{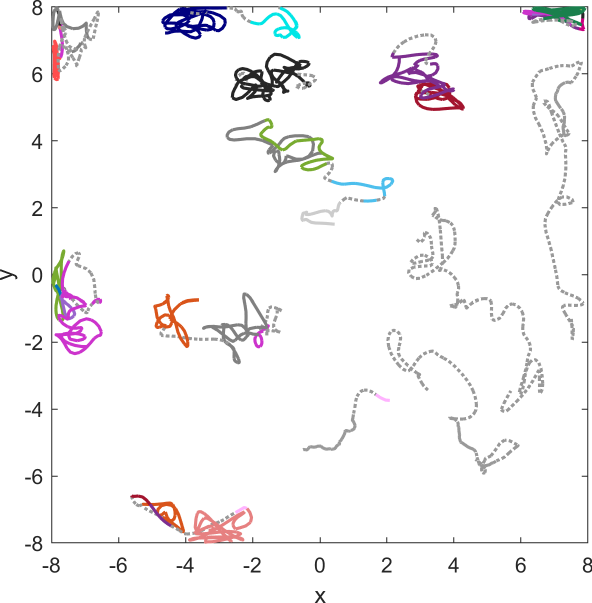}
\caption{Top-on view of particle trajectories with either large or small $\langle S_{\rm max}({\bm X}_{i}(t)\rangle_{\mathbb{T}}$. Different colors represent different coherent sets. Gray dotted sections of trajectories represent times spent in the mixing region. The trajectories were selected such that there is no overlap of different trajectories in this two-dimensional image.}
\label{fig:trajectory3d}
\end{figure}

\begin{figure}[htb]
\centering
\includegraphics[width=0.36\textwidth]{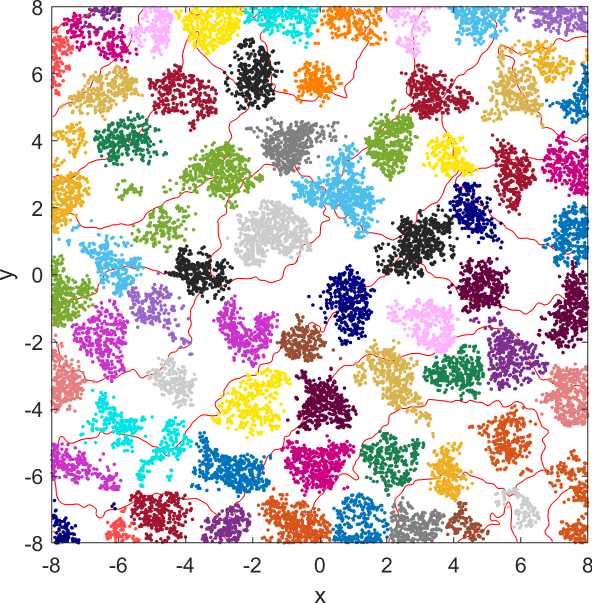}
\caption{Lagrangian clusters at time t, determined by $S_{\rm max}({\bm X}_{i}(t)) \geq 0.7$,  with temperature contour lines $T=0.5$ at mid-plane averaged over $\mathbb{T}_{t,\tau}=\{t-\frac{\tau}{2}, \ldots,  t+\frac{\tau}{2}\}$. The temperature contours indicate the centers of the convection rolls in the Eulerian frame of reference.}
\label{fig:correlation3d}
\end{figure}

Sample trajectories viewed from the top are shown in Fig.~\ref{fig:trajectory3d}. We indicate their membership in a coherent set by solid colored lines and their membership in the spatial complement of the coherent sets by dashed lines. One can clearly see how the individual tracers switch between different coherent sets in the course of their dynamical evolution. The correlation of clusters and temperature contour lines is displayed in Fig.~\ref{fig:correlation3d} which provides a view from the top onto the 3d convection flow. All Lagrangian coherent sets are illustrated together with isocontours of $T=0.5$ in the mid-plane at $z=0.5$. These figures illustrate the increased complexity of the 3d case.

\begin{figure}
\begin{center}
\includegraphics[width=0.5\textwidth]{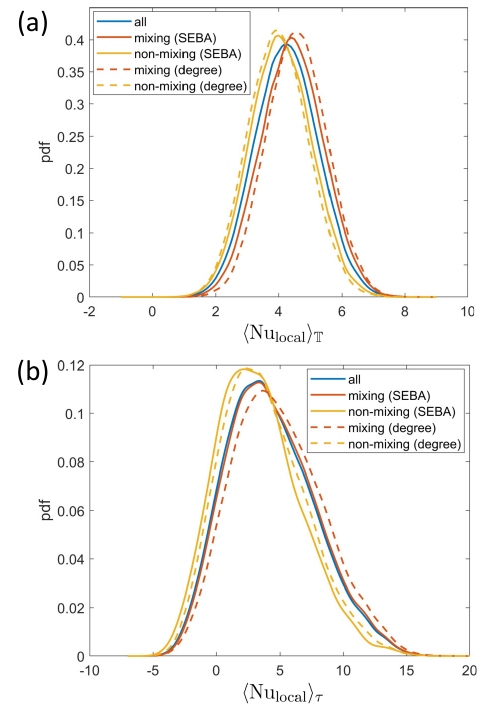}
\caption{\label{fig:results3d}  Heat transport statistics in 3d. (a) pdf of $\langle \Nu_{\rm local}\rangle_{\mathbb{T}}$. Line styles are analogous to those in Fig.~\ref{fig:results2d}(d). (b) The same analysis as in panel (a), but now for $\langle \Nu_{\rm local}\rangle_{\tau}$. All pdfs are smoothed by a Gaussian kernel.}
\end{center}
\end{figure}

\begin{figure}
\begin{center}
\includegraphics[width=0.5\textwidth]{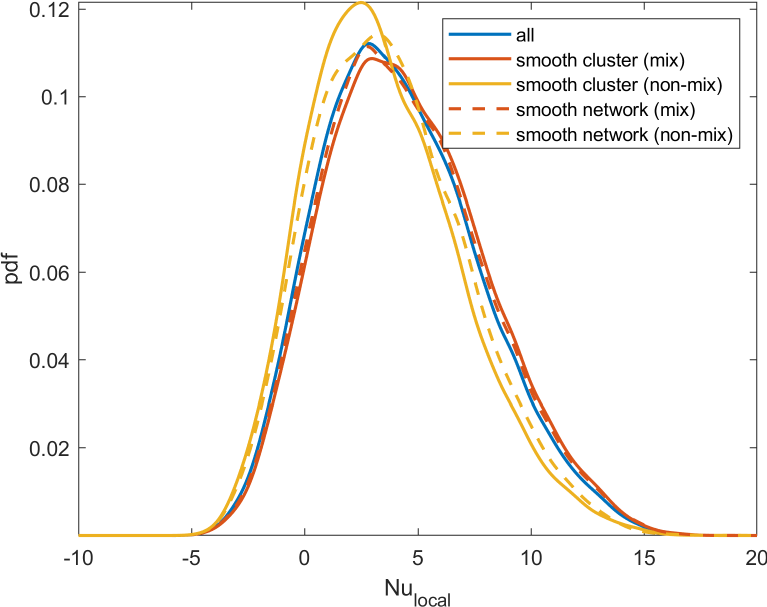}
\caption{\label{fig:approach3d}  Comparison of the heat transport statistics using the evolving cluster approach (solid lines) and the evolving network approach (dashed lines). Red lines correspond to the trajectories with $S_{\rm max}({\bm X}_{i}(t)) < 0.7$, yellow lines to the trajectories with $S_{\rm max}({\bm X}_{i}(t)) \geq 0.7$. All pdfs are smoothed by a Gaussian kernel.}
\end{center}
\end{figure}

Finally, we repeat the analysis of the conditional heat transport along the trajectories. The Nusselt number is now $\langle \Nu_{\rm local}\rangle_{\mathbb{T},L} = 4.3 \pm 1.0$. The corresponding pdfs  are plotted in Fig.~\ref{fig:results3d}(a). Also shown are again the contributions by trajectories assigned to coherent and complementary sets. Based on the cluster membership certainty $S_{\rm max}$, we obtain $\langle \Nu_{\rm local, c}\rangle_{\mathbb{T}} =4.0\pm 1.0$ and $\langle \Nu_{\rm local, m}\rangle_{\mathbb{T}} =4.5 \pm 1.0$. Based on the normalized node degree $d$ it follows $\langle \Nu_{\rm local, c}\rangle_{\mathbb{T}} = 4.0 \pm 1.0$ and $\langle \Nu_{\rm local, m}\rangle_{\mathbb{T}} =4.6 \pm 1.0$. These numbers support the conclusions drawn from the 2d analysis, see also Ref.~\onlinecite{Vieweg2021}. They also demonstrate the applicability of evolutionary clustering to turbulent 3d flows. In addition, we study the heat transport for smaller time intervals. Figure \ref{fig:results3d}(b) shows therefore the pdfs of a single interval $\tau$. The separation of the heat transport contributions are less pronounced in comparison to the full time period $\mathbb{T}$. Nevertheless, the distinction of contributions from coherent sets and their complements is still observable with $\langle \Nu_{\rm local, c}\rangle_{\tau} = 3.3 \pm 3.3$ based on $S_{\rm max}$ compared to $\langle \Nu_{\rm local}\rangle_{\tau, L}= 4.2 \pm 3.5$ for all particles. By choosing a very high threshold value on $S_{\rm max}$ ($S_{\rm max} \geq 0.9$) one can improve considerably the separation into regions with different contributions to heat transport. However, the coherent sets in this case are very diffuse and monitoring the evolution of these sets is difficult. The effect of the evolving cluster approach in contrast to the evolving network approach for the analysis of heat transport contributions is illustrated in Fig.~\ref{fig:approach3d} that shows for one time span $\tau$ the pdf of the instantaneous local Nusselt number. The contributions of the respective distributions for the evolving cluster approach are the red and yellow solid lines. Dashed lines correspond to the equivalent distributions using the evolving network approach. One can see that for this time span the separation into sets with different contributions to heat transport works better using the evolving cluster approach. This is the case for almost all time intervals and also for the analysis of the time averaged local Nusselt number $\langle \Nu_{\rm local}\rangle_{\mathbb{T},L}$ (not shown).

\section{Concluding remarks}
Studies of turbulent convection in the Eulerian frame of reference identified the formation of slowly evolving patterns termed turbulent superstructures. They are formed by rising and falling thermal plumes which subsequently create large-scale convection rolls. The detection of superstructures and the analysis of their long-term evolution in the Lagrangian frame of reference, however, is limited when conventional spectral clustering is applied. The reason is the immanent dispersion in turbulent flows that destroys coherence of material transport for longer time intervals. In the present work, we applied two evolutionary trajectory-based network frameworks to overcome this constraint and to follow detected coherent sets in 2d and 3d turbulent convection flows in their gradual time evolution. This approach eases the mathematical rigor of material transport on the one hand. It allows, on the other hand, to monitor a long-term dynamics of coherent sets including their splits or mergers. These events were traced back to the changes of spectral properties of the corresponding network matrices. Thus, we were able to detect long living sets with lifetimes of more than $10^3$ $T_f$ in the 2d case. We showed that Lagrangian coherent sets have a different dynamical relevance compared to their spatial complements, the strong mixing regions. They effectively suppress the turbulent heat transport across the domain. Their contribution to the turbulent heat transfer is reduced by about 30\% in the two-dimensional and 13\% in the three-dimensional case, respectively. 

Particularly in the three-dimensional turbulent flow case, the application of evolutionary spectral clustering as introduced in section \ref{sec:eca} stabilizes the coherent sets and leads to their identification as sets with minimal contribution to the heat transfer. This is different to the successive spectral clustering. Furthermore, it allows us to study the dynamical behavior of these macroscopic flow structures, which has not been possible by simple time-extension of the spectral clustering approach due to turbulent dispersion \cite{Schneide2018}.
These persistent evolving sets are thus good candidates to serve as the Lagrangian counterpart of the Eulerian turbulent superstructures in turbulent convection flows. The continued investigation of the robustness of this approach as a function of the Rayleigh number for two- and three-dimensional cases as well as with respect to the number of tracer particles is part of our future work. The latter point would be particularly relevant for investigations of the robustness of the method for sparse data records. The presented trajectory-based dynamical framework might also be well-suited when several transport processes with different material parameters have to be considered, such as temperature and salinity in the ocean which differ in their diffusion constants by a factor of $\sim 10^2$. These measurements are often carried out by drifters and the proposed clustering approach can be adapted to deal with such sparse and 
possibly incomplete multivariate data sets.

%

\begin{acknowledgments}
 CS and PPV are supported by the Priority Programme DFG-SPP 1881 ``Turbulent Superstructures'' of the Deutsche Forschungsgemeinschaft. The authors gratefully acknowledge the Gauss Centre for Supercomputing e.V. (www.gauss-centre.eu) for funding this project by providing computing time through the John von Neumann Institute for Computing (NIC) on the GCS Supercomputer JUWELS at J\"ulich Supercomputing Centre (JSC). KPG thanks Alexandra von Kameke for fruitful discussions on the weighted network construction.
\end{acknowledgments}

\section*{Data Availability Statement}
The data that support the findings of this study are available from the corresponding author upon reasonable request.

\appendix
\section{\label{sec:AppSEBA}Sparse Eigenbasis Approximation and evolutionary clustering}
The main idea of the sparse eigenbasis approximation (SEBA) \cite{Froyland2019} is to transform the set of orthonormal eigenvectors $u_1, \ldots, u_k$, which span a subspace $\mathcal{U} \subset \mathbb{R}^{N}$ to a basis of sparse vectors $s_1, \ldots, s_k$ for a subspace $\mathcal{S} \subset \mathbb{R}^{N}$ which approximates $\mathcal{U}$, i.e. $\mathcal{U} \approx \mathcal{S}$. To be more precise, given the matrix $U \in \mathbb{R}^{k \times N}$, the goal is to find a rotation matrix $R \in \mathbb{R}^{k \times k}$ and a sparse matrix $S \in \mathbb{R}^{N \times k}$ such that $U \approx SR$, see Ref.~\onlinecite{Froyland2019} for the exact setting. It immediately follows that 
\begin{equation}
UU^T \approx SR(SR)^T =SRR^TS^T=SS^T.\label{eq:seba}
\end{equation}

In the trajectory clustering framework, different flow components, i.e. regions in the flow with different dynamical behavior, originally encoded in the eigenvectors of the network matrix $D^{-1/2}WD^{-1/2}$, are approximated by individual sparse vectors. The algorithm in Ref.~\onlinecite{Froyland2019} returns the sparse vector $s_i$ as column $i$ of a matrix $S$ where the columns are ordered by decreasing reliability of the identified flow component. Due to a rescaling used in Ref.\,\onlinecite{Froyland2019}, the entry $S_{ij}$ can be interpreted as the likelihood of the data point connected to index $i$ to be an element of component $j$. Therefore, $S_{\rm max, i} = \max_j S_{ij}$ represents the maximum likelihood of data point $i$ to belong to any flow component. By choosing a threshold on $S_{\rm max}$, data points can be assigned to coherent regions or to the complement, the incoherent mixing region. 

In the framework of preserving cluster memberships \cite{Chi2007}, the optimal solution to the normalized cut problem for time $t$ is the matrix $X_t$ whose columns are the eigenvectors of matrix 
$$
\widehat{W}_t=\alpha D_t^{-1/2}W_t D_t^{-1/2} + (1-\alpha) X_{t-1}X_{t-1}^T,
$$ 
where $X_{t-1}$ is obtained in the previous time step.
In order to speed up the computation, we calculate a sparse approximation of $\widehat{W}_t$.
First, we approximate the subspace spanned by $X_{t-1}$ in terms of the corresponding sparse subspace spanned by the columns of the matrix $\tilde{S}_{t-1}$.  $\tilde{S}_{t-1}$ is obtained via the SEBA output matrix $S_{t-1}$ from $X_{t-1}$, setting possibly occurring negative entries in $S_{t-1}$ to zero, and a subsequent column normalization of $S_{t-1}$. Using \eqref{eq:seba} this results in
\begin{equation}
    X_{t-1}X_{t-1}^T \approx \tilde{S}_{t-1}\tilde{S}_{t-1}^T.
\end{equation}
Second, we insert this approximation into the derivation of $\widehat{W}_t$
\begin{equation}
    \widehat{W}_t \approx \alpha D_t^{-1/2}W_t D_t^{-1/2} + (1-\alpha) \tilde{S}_{t-1}\tilde{S}_{t-1}^T =: \tilde{W}_t.
\end{equation}
An approximation of the optimal solution for time $t$ is now given by the matrix whose columns are the eigenvectors of the normalized matrix $\tilde{D}_t^{-1/2}\tilde{W}_t \tilde{D}_t^{-1/2}$, where $\tilde{D}_t$ is the degree matrix of $\tilde{W}_t$.  By this normalization $\tilde{D}_t^{-1/2}\tilde{W}_t \tilde{D}_t^{-1/2}$ is a stochastic matrix.

\section{\label{sec:AppSpectrum}Number of relevant eigenvectors and temporal evolution of eigenvalues}
Spectral clustering makes use of the eigenvectors corresponding to the $k$ leading eigenvalues of the network matrix, where $k$ is heuristically chosen according to a spectral gap criterion. To this end, we compute 
the $K$ largest eigenvalues $\lambda_{t,1}=1 > \lambda_{t,2} \geq \ldots \geq \lambda_{t,K}$ for the network with respect to the time window centered at $t \in \mathbb{T}$. The number $K$ is significantly larger than the expected $k_t$, where $k_t$ denotes the number of relevant eigenvectors for this time step. In this work, we use $K=10$ for the 2d case and $K=100$ for the 3d case. For each time step $t$, we calculate the distances between successive eigenvalues via 
\begin{equation}
 g_t(i)=\lambda_{t,i} - \lambda_{t,i+1}, \; i=1, \ldots, K-1. \end{equation}
and determine the arguments of the (local) maxima of $g_t$. 
For the 2d case, we initially choose $k_t=8$ as this maximizes $g_t$ for the initial time window centered at $t=160 T_f$. 
Under the time evolving network $k_t$ is chosen at each time step to be the maximizer of $g_t$ when $k_t \in \{k_{t-1}-1, k_{t-1}, k_{t-1}+1\}$. 
In the 3d case, we start with $k_t=70$ as a local maximizer of $g_t$ for the initial time window centered at $t=75.3 T_f$, taking into account prior knowledge of the approximate number of clusters from Ref.~\onlinecite{Vieweg2021}. As the evolving cluster approach stabilizes the number of clusters, we choose $k_t \in [0.9k_{t-1}, 1.1k_{t-1}]$ again as a local maximizer of $g_t$, allowing us to monitor significant changes in the time-dependent clusterings. Further heuristics on the spectral-based choice of the number of clusters are discussed in Ref.~\onlinecite{Froyland2019}.

We note that for the sake of a simple and clear visualization of the temporal behavior of the spectrum we have just connected the corresponding (ordered) eigenvalues $\lambda_{t,i}$ and $\lambda_{t+1,i}$ from successive time windows in Figures \ref{fig:evolution2d}(c) and \ref{fig:evolution3d}(c). This is correct as long as these curves are well separated. In order to accurately follow how the eigenvalues evolve in time in the case of crossings (i.e. changes in the ordering of the eigenvalues), one would have to take into account the evolution of the corresponding eigenvectors as proposed in Ref.~\onlinecite{Blachut2020}. However, as we do not consider the temporal behavior of individual eigenvectors in our study, crossings of eigenvalues are difficult to interpret, and moreover, such an investigation is unfeasible for the 3d case where a very large number of eigenvalues and corresponding eigenvectors is studied. 

\bibliography{references}

\begin{thebibliography}{45}%
\makeatletter
\providecommand \@ifxundefined [1]{%
 \@ifx{#1\undefined}
}%
\providecommand \@ifnum [1]{%
 \ifnum #1\expandafter \@firstoftwo
 \else \expandafter \@secondoftwo
 \fi
}%
\providecommand \@ifx [1]{%
 \ifx #1\expandafter \@firstoftwo
 \else \expandafter \@secondoftwo
 \fi
}%
\providecommand \natexlab [1]{#1}%
\providecommand \enquote  [1]{``#1''}%
\providecommand \bibnamefont  [1]{#1}%
\providecommand \bibfnamefont [1]{#1}%
\providecommand \citenamefont [1]{#1}%
\providecommand \href@noop [0]{\@secondoftwo}%
\providecommand \href [0]{\begingroup \@sanitize@url \@href}%
\providecommand \@href[1]{\@@startlink{#1}\@@href}%
\providecommand \@@href[1]{\endgroup#1\@@endlink}%
\providecommand \@sanitize@url [0]{\catcode `\\12\catcode `\$12\catcode
  `\&12\catcode `\#12\catcode `\^12\catcode `\_12\catcode `\%12\relax}%
\providecommand \@@startlink[1]{}%
\providecommand \@@endlink[0]{}%
\providecommand \url  [0]{\begingroup\@sanitize@url \@url }%
\providecommand \@url [1]{\endgroup\@href {#1}{\urlprefix }}%
\providecommand \urlprefix  [0]{URL }%
\providecommand \Eprint [0]{\href }%
\providecommand \doibase [0]{http://dx.doi.org/}%
\providecommand \selectlanguage [0]{\@gobble}%
\providecommand \bibinfo  [0]{\@secondoftwo}%
\providecommand \bibfield  [0]{\@secondoftwo}%
\providecommand \translation [1]{[#1]}%
\providecommand \BibitemOpen [0]{}%
\providecommand \bibitemStop [0]{}%
\providecommand \bibitemNoStop [0]{.\EOS\space}%
\providecommand \EOS [0]{\spacefactor3000\relax}%
\providecommand \BibitemShut  [1]{\csname bibitem#1\endcsname}%
\let\auto@bib@innerbib\@empty
\bibitem [{\citenamefont {Ahlers}, \citenamefont {Grossmann},\ and\
  \citenamefont {Lohse}(2009)}]{Ahlers2009}%
  \BibitemOpen
  \bibfield  {author} {\bibinfo {author} {\bibfnamefont {G.}~\bibnamefont
  {Ahlers}}, \bibinfo {author} {\bibfnamefont {S.}~\bibnamefont {Grossmann}}, \
  and\ \bibinfo {author} {\bibfnamefont {D.}~\bibnamefont {Lohse}},\ }\bibfield
   {title} {\enquote {\bibinfo {title} {Heat transfer and large scale dynamics
  in turbulent {R}ayleigh-{B}\'enard convection},}\ }\href {\doibase
  10.1103/RevModPhys.81.503} {\bibfield  {journal} {\bibinfo  {journal} {Rev.
  Mod. Phys.}\ }\textbf {\bibinfo {volume} {81}},\ \bibinfo {pages} {503--537}
  (\bibinfo {year} {2009})}\BibitemShut {NoStop}%
\bibitem [{\citenamefont {Chill{\`a}}\ and\ \citenamefont
  {Schumacher}(2012)}]{Chilla2012}%
  \BibitemOpen
  \bibfield  {author} {\bibinfo {author} {\bibfnamefont {F.}~\bibnamefont
  {Chill{\`a}}}\ and\ \bibinfo {author} {\bibfnamefont {J.}~\bibnamefont
  {Schumacher}},\ }\bibfield  {title} {\enquote {\bibinfo {title} {New
  perspectives in turbulent {R}ayleigh-{B}{\'e}nard convection},}\ }\href
  {\doibase 10.1140/epje/i2012-12058-1} {\bibfield  {journal} {\bibinfo
  {journal} {Eur. Phys. J. E}\ }\textbf {\bibinfo {volume} {35}},\ \bibinfo
  {pages} {58} (\bibinfo {year} {2012})}\BibitemShut {NoStop}%
\bibitem [{\citenamefont {Stevens}(2005)}]{Stevens2005}%
  \BibitemOpen
  \bibfield  {author} {\bibinfo {author} {\bibfnamefont {B.}~\bibnamefont
  {Stevens}},\ }\bibfield  {title} {\enquote {\bibinfo {title} {Atmospheric
  moist convection},}\ }\href {\doibase 10.1146/annurev.earth.33.092203.122658}
  {\bibfield  {journal} {\bibinfo  {journal} {Annu. Rev. Earth Planet. Sci.}\
  }\textbf {\bibinfo {volume} {33}},\ \bibinfo {pages} {605--643} (\bibinfo
  {year} {2005})}\BibitemShut {NoStop}%
\bibitem [{\citenamefont {Schumacher}\ and\ \citenamefont
  {Sreenivasan}(2020)}]{Schumacher2020}%
  \BibitemOpen
  \bibfield  {author} {\bibinfo {author} {\bibfnamefont {J.}~\bibnamefont
  {Schumacher}}\ and\ \bibinfo {author} {\bibfnamefont {K.~R.}\ \bibnamefont
  {Sreenivasan}},\ }\bibfield  {title} {\enquote {\bibinfo {title} {Colloquium:
  Unusual dynamics of convection in the sun},}\ }\href {\doibase
  10.1103/RevModPhys.92.041001} {\bibfield  {journal} {\bibinfo  {journal}
  {Rev. Mod. Phys.}\ }\textbf {\bibinfo {volume} {92}},\ \bibinfo {pages}
  {041001} (\bibinfo {year} {2020})}\BibitemShut {NoStop}%
\bibitem [{\citenamefont {Kelley}\ and\ \citenamefont
  {Weier}(2018)}]{Kelley2018}%
  \BibitemOpen
  \bibfield  {author} {\bibinfo {author} {\bibfnamefont {D.~H.}\ \bibnamefont
  {Kelley}}\ and\ \bibinfo {author} {\bibfnamefont {T.}~\bibnamefont {Weier}},\
  }\bibfield  {title} {\enquote {\bibinfo {title} {Fluid mechanics of liquid
  metal batteries},}\ }\href {\doibase 10.1115/1.4038699} {\bibfield  {journal}
  {\bibinfo  {journal} {Appl. Mech. Rev.}\ }\textbf {\bibinfo {volume} {70}},\
  \bibinfo {pages} {020801} (\bibinfo {year} {2018})}\BibitemShut {NoStop}%
\bibitem [{\citenamefont {Allshouse}\ and\ \citenamefont
  {Peacock}(2015)}]{Allshouse2015}%
  \BibitemOpen
  \bibfield  {author} {\bibinfo {author} {\bibfnamefont {M.~R.}\ \bibnamefont
  {Allshouse}}\ and\ \bibinfo {author} {\bibfnamefont {T.}~\bibnamefont
  {Peacock}},\ }\bibfield  {title} {\enquote {\bibinfo {title} {Lagrangian
  based methods for coherent structure detection},}\ }\href {\doibase
  http://dx.doi.org/10.1063/1.4922968} {\bibfield  {journal} {\bibinfo
  {journal} {Chaos}\ }\textbf {\bibinfo {volume} {25}},\ \bibinfo {eid}
  {097617} (\bibinfo {year} {2015})}\BibitemShut {NoStop}%
\bibitem [{\citenamefont {Haller}(2015)}]{Haller2015}%
  \BibitemOpen
  \bibfield  {author} {\bibinfo {author} {\bibfnamefont {G.}~\bibnamefont
  {Haller}},\ }\bibfield  {title} {\enquote {\bibinfo {title} {Lagrangian
  coherent structures},}\ }\href {\doibase 10.1146/annurev-fluid-010313-141322}
  {\bibfield  {journal} {\bibinfo  {journal} {Annu. Rev. Fluid Mech.}\ }\textbf
  {\bibinfo {volume} {47}},\ \bibinfo {pages} {137--162} (\bibinfo {year}
  {2015})}\BibitemShut {NoStop}%
\bibitem [{\citenamefont {Hadjighasem}\ \emph {et~al.}(2017)\citenamefont
  {Hadjighasem}, \citenamefont {Farazmand}, \citenamefont {Blazevski},
  \citenamefont {Froyland},\ and\ \citenamefont {Haller}}]{Hadjighasem2017}%
  \BibitemOpen
  \bibfield  {author} {\bibinfo {author} {\bibfnamefont {A.}~\bibnamefont
  {Hadjighasem}}, \bibinfo {author} {\bibfnamefont {M.}~\bibnamefont
  {Farazmand}}, \bibinfo {author} {\bibfnamefont {D.}~\bibnamefont
  {Blazevski}}, \bibinfo {author} {\bibfnamefont {G.}~\bibnamefont {Froyland}},
  \ and\ \bibinfo {author} {\bibfnamefont {G.}~\bibnamefont {Haller}},\
  }\bibfield  {title} {\enquote {\bibinfo {title} {A critical comparison of
  {Lagrangian} methods for coherent structure detection},}\ }\href@noop {}
  {\bibfield  {journal} {\bibinfo  {journal} {Chaos}\ }\textbf {\bibinfo
  {volume} {27}},\ \bibinfo {pages} {053104} (\bibinfo {year}
  {2017})}\BibitemShut {NoStop}%
\bibitem [{\citenamefont {Tew~Kai}\ \emph {et~al.}(2009)\citenamefont
  {Tew~Kai}, \citenamefont {Rossi}, \citenamefont {Sudre}, \citenamefont
  {Weimerskirch}, \citenamefont {Lopez}, \citenamefont {Hernandez-Garcia},
  \citenamefont {Marsac},\ and\ \citenamefont {Gar{\c c}on}}]{Tew2009}%
  \BibitemOpen
  \bibfield  {author} {\bibinfo {author} {\bibfnamefont {E.}~\bibnamefont
  {Tew~Kai}}, \bibinfo {author} {\bibfnamefont {V.}~\bibnamefont {Rossi}},
  \bibinfo {author} {\bibfnamefont {J.}~\bibnamefont {Sudre}}, \bibinfo
  {author} {\bibfnamefont {H.}~\bibnamefont {Weimerskirch}}, \bibinfo {author}
  {\bibfnamefont {C.}~\bibnamefont {Lopez}}, \bibinfo {author} {\bibfnamefont
  {E.}~\bibnamefont {Hernandez-Garcia}}, \bibinfo {author} {\bibfnamefont
  {F.}~\bibnamefont {Marsac}}, \ and\ \bibinfo {author} {\bibfnamefont
  {V.}~\bibnamefont {Gar{\c c}on}},\ }\bibfield  {title} {\enquote {\bibinfo
  {title} {Top marine predators track {L}agrangian coherent structures},}\
  }\href {\doibase 10.1073/pnas.0811034106} {\bibfield  {journal} {\bibinfo
  {journal} {Proc. Natl. Acad. Sci. USA}\ }\textbf {\bibinfo {volume} {106}},\
  \bibinfo {pages} {8245--8250} (\bibinfo {year} {2009})}\BibitemShut {NoStop}%
\bibitem [{\citenamefont {Shadden}\ and\ \citenamefont
  {Taylor}(2008)}]{Shadden2008}%
  \BibitemOpen
  \bibfield  {author} {\bibinfo {author} {\bibfnamefont {S.}~\bibnamefont
  {Shadden}}\ and\ \bibinfo {author} {\bibfnamefont {C.}~\bibnamefont
  {Taylor}},\ }\bibfield  {title} {\enquote {\bibinfo {title} {Characterization
  of coherent structures in the cardiovascular system},}\ }\href {\doibase
  10.1007/s10439-008-9502-3} {\bibfield  {journal} {\bibinfo  {journal} {Ann.
  Biomed. Eng.}\ }\textbf {\bibinfo {volume} {36}},\ \bibinfo {pages}
  {1152--1162} (\bibinfo {year} {2008})}\BibitemShut {NoStop}%
\bibitem [{\citenamefont {Gawlik}\ \emph {et~al.}(2009)\citenamefont {Gawlik},
  \citenamefont {Marsden}, \citenamefont {Toit},\ and\ \citenamefont
  {Campagnola}}]{Gawlik2009}%
  \BibitemOpen
  \bibfield  {author} {\bibinfo {author} {\bibfnamefont {E.~S.}\ \bibnamefont
  {Gawlik}}, \bibinfo {author} {\bibfnamefont {J.~E.}\ \bibnamefont {Marsden}},
  \bibinfo {author} {\bibfnamefont {P.~C.~D.}\ \bibnamefont {Toit}}, \ and\
  \bibinfo {author} {\bibfnamefont {S.}~\bibnamefont {Campagnola}},\ }\bibfield
   {title} {\enquote {\bibinfo {title} {Lagrangian coherent structures in the
  planar elliptic restricted three-body problem},}\ }\href {\doibase
  10.1007/s10569-008-9180-3} {\bibfield  {journal} {\bibinfo  {journal}
  {Celest. Mech. Dyn. Astron.}\ }\textbf {\bibinfo {volume} {103}},\ \bibinfo
  {pages} {227--249} (\bibinfo {year} {2009})}\BibitemShut {NoStop}%
\bibitem [{\citenamefont {Olascoaga}\ and\ \citenamefont
  {Haller}(2012)}]{Olascoaga2012}%
  \BibitemOpen
  \bibfield  {author} {\bibinfo {author} {\bibfnamefont {M.~J.}\ \bibnamefont
  {Olascoaga}}\ and\ \bibinfo {author} {\bibfnamefont {G.}~\bibnamefont
  {Haller}},\ }\bibfield  {title} {\enquote {\bibinfo {title} {Forecasting
  sudden changes in environmental pollution patterns},}\ }\href {\doibase
  10.1073/pnas.1118574109} {\bibfield  {journal} {\bibinfo  {journal} {Proc.
  Natl. Acad. Sci. USA}\ }\textbf {\bibinfo {volume} {109}},\ \bibinfo {pages}
  {4738--4743} (\bibinfo {year} {2012})}\BibitemShut {NoStop}%
\bibitem [{\citenamefont {Zambianchi}, \citenamefont {Trani},\ and\
  \citenamefont {Falco}(2017)}]{Zambianchi2017}%
  \BibitemOpen
  \bibfield  {author} {\bibinfo {author} {\bibfnamefont {E.}~\bibnamefont
  {Zambianchi}}, \bibinfo {author} {\bibfnamefont {M.}~\bibnamefont {Trani}}, \
  and\ \bibinfo {author} {\bibfnamefont {P.}~\bibnamefont {Falco}},\ }\bibfield
   {title} {\enquote {\bibinfo {title} {Lagrangian transport of marine litter
  in the mediterranean sea},}\ }\href {\doibase 10.3389/fenvs.2017.00005}
  {\bibfield  {journal} {\bibinfo  {journal} {Front. Environ. Sci.}\ }\textbf
  {\bibinfo {volume} {5}},\ \bibinfo {pages} {5} (\bibinfo {year}
  {2017})}\BibitemShut {NoStop}%
\bibitem [{\citenamefont {Froyland}, \citenamefont {Lloyd},\ and\ \citenamefont
  {Santitissadeekorn}(2010)}]{FroylandLloydSan2010}%
  \BibitemOpen
  \bibfield  {author} {\bibinfo {author} {\bibfnamefont {G.}~\bibnamefont
  {Froyland}}, \bibinfo {author} {\bibfnamefont {S.}~\bibnamefont {Lloyd}}, \
  and\ \bibinfo {author} {\bibfnamefont {N.}~\bibnamefont
  {Santitissadeekorn}},\ }\bibfield  {title} {\enquote {\bibinfo {title}
  {Coherent sets for nonautonomous dynamical systems},}\ }\href@noop {}
  {\bibfield  {journal} {\bibinfo  {journal} {Physica D}\ }\textbf {\bibinfo
  {volume} {239}},\ \bibinfo {pages} {1527--1541} (\bibinfo {year}
  {2010})}\BibitemShut {NoStop}%
\bibitem [{\citenamefont {Froyland}(2013)}]{Froyland2013}%
  \BibitemOpen
  \bibfield  {author} {\bibinfo {author} {\bibfnamefont {G.}~\bibnamefont
  {Froyland}},\ }\bibfield  {title} {\enquote {\bibinfo {title} {An analytic
  framework for identifying finite-time coherent sets in time-dependent
  dynamical systems},}\ }\href@noop {} {\bibfield  {journal} {\bibinfo
  {journal} {Physica D}\ }\textbf {\bibinfo {volume} {250}},\ \bibinfo {pages}
  {1--19} (\bibinfo {year} {2013})}\BibitemShut {NoStop}%
\bibitem [{\citenamefont {Karrasch}\ and\ \citenamefont
  {Keller}(2020)}]{Karrasch2020}%
  \BibitemOpen
  \bibfield  {author} {\bibinfo {author} {\bibfnamefont {D.}~\bibnamefont
  {Karrasch}}\ and\ \bibinfo {author} {\bibfnamefont {J.}~\bibnamefont
  {Keller}},\ }\bibfield  {title} {\enquote {\bibinfo {title} {A geometric
  heat-flow theory of {L}agrangian coherent structures},}\ }\href {\doibase
  10.1007/s00332-020-09626-9} {\bibfield  {journal} {\bibinfo  {journal} {J.
  Nonlinear Sci.}\ }\textbf {\bibinfo {volume} {30}},\ \bibinfo {pages}
  {1849--1888} (\bibinfo {year} {2020})}\BibitemShut {NoStop}%
\bibitem [{\citenamefont {Haller}, \citenamefont {Karrasch},\ and\
  \citenamefont {Kogelbauer}(2018)}]{Haller2018}%
  \BibitemOpen
  \bibfield  {author} {\bibinfo {author} {\bibfnamefont {G.}~\bibnamefont
  {Haller}}, \bibinfo {author} {\bibfnamefont {D.}~\bibnamefont {Karrasch}}, \
  and\ \bibinfo {author} {\bibfnamefont {F.}~\bibnamefont {Kogelbauer}},\
  }\bibfield  {title} {\enquote {\bibinfo {title} {Material barriers to
  diffusive and stochastic transport},}\ }\href {\doibase
  10.1073/pnas.1720177115} {\bibfield  {journal} {\bibinfo  {journal} {Proc.
  Natl. Acad. Sci. USA}\ }\textbf {\bibinfo {volume} {115}},\ \bibinfo {pages}
  {9074--9079} (\bibinfo {year} {2018})}\BibitemShut {NoStop}%
\bibitem [{\citenamefont {Balasuriya}, \citenamefont {Ouellette},\ and\
  \citenamefont {Rypina}(2018)}]{Balasuriya2018}%
  \BibitemOpen
  \bibfield  {author} {\bibinfo {author} {\bibfnamefont {S.}~\bibnamefont
  {Balasuriya}}, \bibinfo {author} {\bibfnamefont {N.~T.}\ \bibnamefont
  {Ouellette}}, \ and\ \bibinfo {author} {\bibfnamefont {I.~I.}\ \bibnamefont
  {Rypina}},\ }\bibfield  {title} {\enquote {\bibinfo {title} {Generalized
  lagrangian coherent structures},}\ }\href {\doibase
  https://doi.org/10.1016/j.physd.2018.01.011} {\bibfield  {journal} {\bibinfo
  {journal} {Physica D: Nonlinear Phenomena}\ }\textbf {\bibinfo {volume}
  {372}},\ \bibinfo {pages} {31--51} (\bibinfo {year} {2018})}\BibitemShut
  {NoStop}%
\bibitem [{\citenamefont {Froyland}(2015)}]{Froyland2015}%
  \BibitemOpen
  \bibfield  {author} {\bibinfo {author} {\bibfnamefont {G.}~\bibnamefont
  {Froyland}},\ }\bibfield  {title} {\enquote {\bibinfo {title} {Dynamic
  isoperimetry and the geometry of {Lagrangian} coherent structures},}\ }\href
  {http://stacks.iop.org/0951-7715/28/i=10/a=3587} {\bibfield  {journal}
  {\bibinfo  {journal} {Nonlinearity}\ }\textbf {\bibinfo {volume} {28}},\
  \bibinfo {pages} {3587} (\bibinfo {year} {2015})}\BibitemShut {NoStop}%
\bibitem [{\citenamefont {Froyland}\ and\ \citenamefont
  {Koltai}(2021)}]{Froyland2021}%
  \BibitemOpen
  \bibfield  {author} {\bibinfo {author} {\bibfnamefont {G.}~\bibnamefont
  {Froyland}}\ and\ \bibinfo {author} {\bibfnamefont {P.}~\bibnamefont
  {Koltai}},\ }\bibfield  {title} {\enquote {\bibinfo {title} {Detecting the
  birth and death of finite-time coherent sets},}\ }\href@noop {} {\bibfield
  {journal} {\bibinfo  {journal} {Preprint}\ } (\bibinfo {year} {2021})},\
  \bibinfo {note} {arXiv:2103.16286 [math.DS]}\BibitemShut {NoStop}%
\bibitem [{\citenamefont {Froyland}\ and\ \citenamefont
  {Padberg-Gehle}(2015)}]{FroylandPadberg2015}%
  \BibitemOpen
  \bibfield  {author} {\bibinfo {author} {\bibfnamefont {G.}~\bibnamefont
  {Froyland}}\ and\ \bibinfo {author} {\bibfnamefont {K.}~\bibnamefont
  {Padberg-Gehle}},\ }\bibfield  {title} {\enquote {\bibinfo {title} {A
  rough-and-ready cluster-based approach for extracting finite-time coherent
  sets from sparse and incomplete trajectory data},}\ }\href {\doibase
  http://dx.doi.org/10.1063/1.4926372} {\bibfield  {journal} {\bibinfo
  {journal} {Chaos}\ }\textbf {\bibinfo {volume} {25}},\ \bibinfo {pages}
  {087406} (\bibinfo {year} {2015})}\BibitemShut {NoStop}%
\bibitem [{\citenamefont {Hadjighasem}\ \emph {et~al.}(2016)\citenamefont
  {Hadjighasem}, \citenamefont {Karrasch}, \citenamefont {Teramoto},\ and\
  \citenamefont {Haller}}]{Hadjighasem2016}%
  \BibitemOpen
  \bibfield  {author} {\bibinfo {author} {\bibfnamefont {A.}~\bibnamefont
  {Hadjighasem}}, \bibinfo {author} {\bibfnamefont {D.}~\bibnamefont
  {Karrasch}}, \bibinfo {author} {\bibfnamefont {H.}~\bibnamefont {Teramoto}},
  \ and\ \bibinfo {author} {\bibfnamefont {G.}~\bibnamefont {Haller}},\
  }\bibfield  {title} {\enquote {\bibinfo {title} {Spectral-clustering approach
  to {L}agrangian vortex detection},}\ }\href@noop {} {\bibfield  {journal}
  {\bibinfo  {journal} {Phys. Rev. E}\ }\textbf {\bibinfo {volume} {93}},\
  \bibinfo {pages} {063107} (\bibinfo {year} {2016})}\BibitemShut {NoStop}%
\bibitem [{\citenamefont {Banisch}\ and\ \citenamefont
  {Koltai}(2017)}]{Banisch2017}%
  \BibitemOpen
  \bibfield  {author} {\bibinfo {author} {\bibfnamefont {R.}~\bibnamefont
  {Banisch}}\ and\ \bibinfo {author} {\bibfnamefont {P.}~\bibnamefont
  {Koltai}},\ }\bibfield  {title} {\enquote {\bibinfo {title} {Understanding
  the geometry of transport: Diffusion maps for {L}agrangian trajectory data
  unravel coherent sets},}\ }\href@noop {} {\bibfield  {journal} {\bibinfo
  {journal} {Chaos}\ }\textbf {\bibinfo {volume} {27}},\ \bibinfo {pages}
  {035804} (\bibinfo {year} {2017})}\BibitemShut {NoStop}%
\bibitem [{\citenamefont {Schlueter-Kuck}\ and\ \citenamefont
  {Dabiri}(2017)}]{Schlueter2017}%
  \BibitemOpen
  \bibfield  {author} {\bibinfo {author} {\bibfnamefont {K.~L.}\ \bibnamefont
  {Schlueter-Kuck}}\ and\ \bibinfo {author} {\bibfnamefont {J.~O.}\
  \bibnamefont {Dabiri}},\ }\bibfield  {title} {\enquote {\bibinfo {title}
  {Coherent structure colouring: identification of coherent structures from
  sparse data using graph theory},}\ }\href@noop {} {\bibfield  {journal}
  {\bibinfo  {journal} {J. Fluid Mech.}\ }\textbf {\bibinfo {volume} {811}},\
  \bibinfo {pages} {468--486} (\bibinfo {year} {2017})}\BibitemShut {NoStop}%
\bibitem [{\citenamefont {Padberg-Gehle}\ and\ \citenamefont
  {Schneide}(2017)}]{Padberg2017}%
  \BibitemOpen
  \bibfield  {author} {\bibinfo {author} {\bibfnamefont {K.}~\bibnamefont
  {Padberg-Gehle}}\ and\ \bibinfo {author} {\bibfnamefont {C.}~\bibnamefont
  {Schneide}},\ }\bibfield  {title} {\enquote {\bibinfo {title} {Network-based
  study of {L}agrangian transport and mixing},}\ }\href {\doibase
  10.5194/npg-24-661-2017} {\bibfield  {journal} {\bibinfo  {journal} {Nonlin.
  Processes Geophys.}\ }\textbf {\bibinfo {volume} {24}},\ \bibinfo {pages}
  {661--671} (\bibinfo {year} {2017})}\BibitemShut {NoStop}%
\bibitem [{\citenamefont {Schneide}\ \emph {et~al.}(2018)\citenamefont
  {Schneide}, \citenamefont {Pandey}, \citenamefont {Padberg-Gehle},\ and\
  \citenamefont {Schumacher}}]{Schneide2018}%
  \BibitemOpen
  \bibfield  {author} {\bibinfo {author} {\bibfnamefont {C.}~\bibnamefont
  {Schneide}}, \bibinfo {author} {\bibfnamefont {A.}~\bibnamefont {Pandey}},
  \bibinfo {author} {\bibfnamefont {K.}~\bibnamefont {Padberg-Gehle}}, \ and\
  \bibinfo {author} {\bibfnamefont {J.}~\bibnamefont {Schumacher}},\ }\bibfield
   {title} {\enquote {\bibinfo {title} {Probing turbulent superstructures in
  {R}ayleigh-{B}\'enard convection by {L}agrangian trajectory clusters},}\
  }\href {\doibase 10.1103/PhysRevFluids.3.113501} {\bibfield  {journal}
  {\bibinfo  {journal} {Phys. Rev. Fluids}\ }\textbf {\bibinfo {volume} {3}},\
  \bibinfo {pages} {113501} (\bibinfo {year} {2018})}\BibitemShut {NoStop}%
\bibitem [{\citenamefont {Vieweg}\ \emph {et~al.}(2021)\citenamefont {Vieweg},
  \citenamefont {Schneide}, \citenamefont {Padberg-Gehle},\ and\ \citenamefont
  {Schumacher}}]{Vieweg2021}%
  \BibitemOpen
  \bibfield  {author} {\bibinfo {author} {\bibfnamefont {P.~P.}\ \bibnamefont
  {Vieweg}}, \bibinfo {author} {\bibfnamefont {C.}~\bibnamefont {Schneide}},
  \bibinfo {author} {\bibfnamefont {K.}~\bibnamefont {Padberg-Gehle}}, \ and\
  \bibinfo {author} {\bibfnamefont {J.}~\bibnamefont {Schumacher}},\ }\bibfield
   {title} {\enquote {\bibinfo {title} {Lagrangian heat transport in turbulent
  three-dimensional convection},}\ }\href {\doibase
  10.1103/PhysRevFluids.6.L041501} {\bibfield  {journal} {\bibinfo  {journal}
  {Phys. Rev. Fluids}\ }\textbf {\bibinfo {volume} {6}},\ \bibinfo {pages}
  {L041501} (\bibinfo {year} {2021})}\BibitemShut {NoStop}%
\bibitem [{\citenamefont {Brunton}, \citenamefont {Noack},\ and\ \citenamefont
  {Koumoutsakos}(2020)}]{Brunton2020}%
  \BibitemOpen
  \bibfield  {author} {\bibinfo {author} {\bibfnamefont {S.~L.}\ \bibnamefont
  {Brunton}}, \bibinfo {author} {\bibfnamefont {B.~R.}\ \bibnamefont {Noack}},
  \ and\ \bibinfo {author} {\bibfnamefont {P.}~\bibnamefont {Koumoutsakos}},\
  }\bibfield  {title} {\enquote {\bibinfo {title} {Machine learning for fluid
  mechanics},}\ }\href {\doibase 10.1146/annurev-fluid-010719-060214}
  {\bibfield  {journal} {\bibinfo  {journal} {Annu. Rev. Fluid Mech.}\ }\textbf
  {\bibinfo {volume} {52}},\ \bibinfo {pages} {477--508} (\bibinfo {year}
  {2020})}\BibitemShut {NoStop}%
\bibitem [{\citenamefont {Stevens}\ \emph {et~al.}(2018)\citenamefont
  {Stevens}, \citenamefont {Blass}, \citenamefont {Zhu}, \citenamefont
  {Verzicco},\ and\ \citenamefont {Lohse}}]{Stevens2018}%
  \BibitemOpen
  \bibfield  {author} {\bibinfo {author} {\bibfnamefont {R.~J. A.~M.}\
  \bibnamefont {Stevens}}, \bibinfo {author} {\bibfnamefont {A.}~\bibnamefont
  {Blass}}, \bibinfo {author} {\bibfnamefont {X.}~\bibnamefont {Zhu}}, \bibinfo
  {author} {\bibfnamefont {R.}~\bibnamefont {Verzicco}}, \ and\ \bibinfo
  {author} {\bibfnamefont {D.}~\bibnamefont {Lohse}},\ }\bibfield  {title}
  {\enquote {\bibinfo {title} {Turbulent thermal superstructures in
  {R}ayleigh-{B}\'enard convection},}\ }\href {\doibase
  10.1103/PhysRevFluids.3.041501} {\bibfield  {journal} {\bibinfo  {journal}
  {Phys. Rev. Fluids}\ }\textbf {\bibinfo {volume} {3}},\ \bibinfo {pages}
  {041501} (\bibinfo {year} {2018})}\BibitemShut {NoStop}%
\bibitem [{\citenamefont {Pandey}, \citenamefont {Scheel},\ and\ \citenamefont
  {Schumacher}(2018)}]{Pandey2018}%
  \BibitemOpen
  \bibfield  {author} {\bibinfo {author} {\bibfnamefont {A.}~\bibnamefont
  {Pandey}}, \bibinfo {author} {\bibfnamefont {J.~D.}\ \bibnamefont {Scheel}},
  \ and\ \bibinfo {author} {\bibfnamefont {J.}~\bibnamefont {Schumacher}},\
  }\bibfield  {title} {\enquote {\bibinfo {title} {Turbulent superstructures in
  {R}ayleigh-{B}{\'e}nard convection},}\ }\href {\doibase
  10.1038/s41467-018-04478-0} {\bibfield  {journal} {\bibinfo  {journal} {Nat.
  Commun.}\ }\textbf {\bibinfo {volume} {9}},\ \bibinfo {pages} {2118}
  (\bibinfo {year} {2018})}\BibitemShut {NoStop}%
\bibitem [{\citenamefont {Green}\ \emph {et~al.}(2020)\citenamefont {Green},
  \citenamefont {Vlaykov}, \citenamefont {Mellado},\ and\ \citenamefont
  {Wilczek}}]{Green2020}%
  \BibitemOpen
  \bibfield  {author} {\bibinfo {author} {\bibfnamefont {G.}~\bibnamefont
  {Green}}, \bibinfo {author} {\bibfnamefont {D.~G.}\ \bibnamefont {Vlaykov}},
  \bibinfo {author} {\bibfnamefont {J.~P.}\ \bibnamefont {Mellado}}, \ and\
  \bibinfo {author} {\bibfnamefont {M.}~\bibnamefont {Wilczek}},\ }\bibfield
  {title} {\enquote {\bibinfo {title} {Resolved energy budget of
  superstructures in {R}ayleigh-{B}\'enard convection},}\ }\href@noop {}
  {\bibfield  {journal} {\bibinfo  {journal} {J. Fluid Mech.}\ }\textbf
  {\bibinfo {volume} {887}},\ \bibinfo {pages} {A21} (\bibinfo {year}
  {2020})}\BibitemShut {NoStop}%
\bibitem [{\citenamefont {Krug}, \citenamefont {Lohse},\ and\ \citenamefont
  {Stevens}(2020)}]{Krug2020}%
  \BibitemOpen
  \bibfield  {author} {\bibinfo {author} {\bibfnamefont {D.}~\bibnamefont
  {Krug}}, \bibinfo {author} {\bibfnamefont {D.}~\bibnamefont {Lohse}}, \ and\
  \bibinfo {author} {\bibfnamefont {R.~J. A.~M.}\ \bibnamefont {Stevens}},\
  }\bibfield  {title} {\enquote {\bibinfo {title} {Coherence of temperature and
  velocity superstructures in turbulent {R}ayleigh-{B\'e}nard flow},}\ }\href
  {\doibase 10.1017/jfm.2019.1054} {\bibfield  {journal} {\bibinfo  {journal}
  {J. Fluid Mech.}\ }\textbf {\bibinfo {volume} {887}},\ \bibinfo {pages} {A2}
  (\bibinfo {year} {2020})}\BibitemShut {NoStop}%
\bibitem [{\citenamefont {Vieweg}, \citenamefont {Scheel},\ and\ \citenamefont
  {Schumacher}(2021)}]{Vieweg2021a}%
  \BibitemOpen
  \bibfield  {author} {\bibinfo {author} {\bibfnamefont {P.~P.}\ \bibnamefont
  {Vieweg}}, \bibinfo {author} {\bibfnamefont {J.~D.}\ \bibnamefont {Scheel}},
  \ and\ \bibinfo {author} {\bibfnamefont {J.}~\bibnamefont {Schumacher}},\
  }\bibfield  {title} {\enquote {\bibinfo {title} {Supergranule aggregation for
  constant heat flux-driven turbulent convection},}\ }\href {\doibase
  10.1103/PhysRevResearch.3.013231} {\bibfield  {journal} {\bibinfo  {journal}
  {Phys. Rev. Res.}\ }\textbf {\bibinfo {volume} {3}},\ \bibinfo {pages}
  {013231} (\bibinfo {year} {2021})}\BibitemShut {NoStop}%
\bibitem [{\citenamefont {Fonda}\ \emph {et~al.}(2019)\citenamefont {Fonda},
  \citenamefont {Pandey}, \citenamefont {Schumacher},\ and\ \citenamefont
  {Sreenivasan}}]{Fonda2019}%
  \BibitemOpen
  \bibfield  {author} {\bibinfo {author} {\bibfnamefont {E.}~\bibnamefont
  {Fonda}}, \bibinfo {author} {\bibfnamefont {A.}~\bibnamefont {Pandey}},
  \bibinfo {author} {\bibfnamefont {J.}~\bibnamefont {Schumacher}}, \ and\
  \bibinfo {author} {\bibfnamefont {K.~R.}\ \bibnamefont {Sreenivasan}},\
  }\bibfield  {title} {\enquote {\bibinfo {title} {Deep learning in turbulent
  convection networks},}\ }\href@noop {} {\bibfield  {journal} {\bibinfo
  {journal} {Proc. Natl. Acad. Sci. USA}\ }\textbf {\bibinfo {volume} {116}},\
  \bibinfo {pages} {8667--8672} (\bibinfo {year} {2019})}\BibitemShut {NoStop}%
\bibitem [{\citenamefont {MacMillan}, \citenamefont {Ouellette},\ and\
  \citenamefont {Richter}(2020)}]{MacMillan2020}%
  \BibitemOpen
  \bibfield  {author} {\bibinfo {author} {\bibfnamefont {T.}~\bibnamefont
  {MacMillan}}, \bibinfo {author} {\bibfnamefont {N.~T.}\ \bibnamefont
  {Ouellette}}, \ and\ \bibinfo {author} {\bibfnamefont {D.~H.}\ \bibnamefont
  {Richter}},\ }\bibfield  {title} {\enquote {\bibinfo {title} {Detection of
  evolving {Lagrangian} coherent structures: A multiple object tracking
  approach},}\ }\href {\doibase 10.1103/PhysRevFluids.5.124401} {\bibfield
  {journal} {\bibinfo  {journal} {Phys. Rev. Fluids}\ }\textbf {\bibinfo
  {volume} {5}},\ \bibinfo {pages} {124401} (\bibinfo {year}
  {2020})}\BibitemShut {NoStop}%
\bibitem [{\citenamefont {Chi}\ \emph {et~al.}(2007)\citenamefont {Chi},
  \citenamefont {Song}, \citenamefont {Zhou}, \citenamefont {Hino},\ and\
  \citenamefont {Tseng}}]{Chi2007}%
  \BibitemOpen
  \bibfield  {author} {\bibinfo {author} {\bibfnamefont {Y.}~\bibnamefont
  {Chi}}, \bibinfo {author} {\bibfnamefont {X.}~\bibnamefont {Song}}, \bibinfo
  {author} {\bibfnamefont {D.}~\bibnamefont {Zhou}}, \bibinfo {author}
  {\bibfnamefont {K.}~\bibnamefont {Hino}}, \ and\ \bibinfo {author}
  {\bibfnamefont {B.~L.}\ \bibnamefont {Tseng}},\ }\bibfield  {title} {\enquote
  {\bibinfo {title} {Evolutionary spectral clustering by incorporating temporal
  smoothness},}\ }in\ \href {\doibase 10.1145/1281192.1281212} {\emph {\bibinfo
  {booktitle} {Proceedings of the 13th ACM SIGKDD International Conference on
  Knowledge Discovery and Data Mining (KDD-07)}}}\ (\bibinfo {address} {New
  York, NY, USA},\ \bibinfo {year} {2007})\ pp.\ \bibinfo {pages}
  {153--162}\BibitemShut {NoStop}%
\bibitem [{\citenamefont {Froyland}, \citenamefont {Rock},\ and\ \citenamefont
  {Sakellariou}(2019)}]{Froyland2019}%
  \BibitemOpen
  \bibfield  {author} {\bibinfo {author} {\bibfnamefont {G.}~\bibnamefont
  {Froyland}}, \bibinfo {author} {\bibfnamefont {C.~P.}\ \bibnamefont {Rock}},
  \ and\ \bibinfo {author} {\bibfnamefont {K.}~\bibnamefont {Sakellariou}},\
  }\bibfield  {title} {\enquote {\bibinfo {title} {Sparse eigenbasis
  approximation: Multiple feature extraction across spatiotemporal scales with
  application to coherent set identification},}\ }\href {\doibase
  https://doi.org/10.1016/j.cnsns.2019.04.012} {\bibfield  {journal} {\bibinfo
  {journal} {Commun. Nonlinear Sci. Numer. Simul.}\ }\textbf {\bibinfo {volume}
  {77}},\ \bibinfo {pages} {81--107} (\bibinfo {year} {2019})}\BibitemShut
  {NoStop}%
\bibitem [{\citenamefont {Scheel}, \citenamefont {Emran},\ and\ \citenamefont
  {Schumacher}(2013)}]{Scheel2013}%
  \BibitemOpen
  \bibfield  {author} {\bibinfo {author} {\bibfnamefont {J.~D.}\ \bibnamefont
  {Scheel}}, \bibinfo {author} {\bibfnamefont {M.~S.}\ \bibnamefont {Emran}}, \
  and\ \bibinfo {author} {\bibfnamefont {J.}~\bibnamefont {Schumacher}},\
  }\bibfield  {title} {\enquote {\bibinfo {title} {Resolving the fine-scale
  structure in turbulent {R}ayleigh-{B}\'enard convection},}\ }\href {\doibase
  10.1088/1367-2630/15/11/113063} {\bibfield  {journal} {\bibinfo  {journal}
  {New J. Phys.}\ }\textbf {\bibinfo {volume} {15}},\ \bibinfo {pages} {113063}
  (\bibinfo {year} {2013})}\BibitemShut {NoStop}%
\bibitem [{\citenamefont {Iacobello}, \citenamefont {Ridolfi},\ and\
  \citenamefont {Scarsoglio}(2021)}]{Iacobello2021}%
  \BibitemOpen
  \bibfield  {author} {\bibinfo {author} {\bibfnamefont {G.}~\bibnamefont
  {Iacobello}}, \bibinfo {author} {\bibfnamefont {L.}~\bibnamefont {Ridolfi}},
  \ and\ \bibinfo {author} {\bibfnamefont {S.}~\bibnamefont {Scarsoglio}},\
  }\bibfield  {title} {\enquote {\bibinfo {title} {A review on turbulent and
  vortical flow analyses via complex networks},}\ }\href {\doibase
  https://doi.org/10.1016/j.physa.2020.125476} {\bibfield  {journal} {\bibinfo
  {journal} {Physica A}\ }\textbf {\bibinfo {volume} {563}},\ \bibinfo {pages}
  {125476} (\bibinfo {year} {2021})}\BibitemShut {NoStop}%
\bibitem [{\citenamefont {Banisch}, \citenamefont {Koltai},\ and\ \citenamefont
  {Padberg-Gehle}(2019)}]{Banisch2019}%
  \BibitemOpen
  \bibfield  {author} {\bibinfo {author} {\bibfnamefont {R.}~\bibnamefont
  {Banisch}}, \bibinfo {author} {\bibfnamefont {P.}~\bibnamefont {Koltai}}, \
  and\ \bibinfo {author} {\bibfnamefont {K.}~\bibnamefont {Padberg-Gehle}},\
  }\bibfield  {title} {\enquote {\bibinfo {title} {Network measures of
  mixing},}\ }\href {\doibase 10.1063/1.5087632} {\bibfield  {journal}
  {\bibinfo  {journal} {Chaos}\ }\textbf {\bibinfo {volume} {29}},\ \bibinfo
  {pages} {063125} (\bibinfo {year} {2019})}\BibitemShut {NoStop}%
\bibitem [{\citenamefont {Shi}\ and\ \citenamefont
  {Malik}(2000)}]{ShiMalik2000}%
  \BibitemOpen
  \bibfield  {author} {\bibinfo {author} {\bibfnamefont {J.}~\bibnamefont
  {Shi}}\ and\ \bibinfo {author} {\bibfnamefont {J.}~\bibnamefont {Malik}},\
  }\bibfield  {title} {\enquote {\bibinfo {title} {Normalized cuts and image
  segmentation},}\ }\href {\doibase 10.1109/34.868688} {\bibfield  {journal}
  {\bibinfo  {journal} {IEEE Trans. Pattern Anal. Mach. Intell.}\ }\textbf
  {\bibinfo {volume} {22}},\ \bibinfo {pages} {888--905} (\bibinfo {year}
  {2000})}\BibitemShut {NoStop}%
\bibitem [{\citenamefont {Bach}\ and\ \citenamefont {Jordan}(2006)}]{Bach2006}%
  \BibitemOpen
  \bibfield  {author} {\bibinfo {author} {\bibfnamefont {F.~R.}\ \bibnamefont
  {Bach}}\ and\ \bibinfo {author} {\bibfnamefont {M.~I.}\ \bibnamefont
  {Jordan}},\ }\bibfield  {title} {\enquote {\bibinfo {title} {Learning
  spectral clustering, with application to speech separation},}\ }\href
  {http://jmlr.org/papers/v7/bach06b.html} {\bibfield  {journal} {\bibinfo
  {journal} {J. Mach. Learn. Res.}\ }\textbf {\bibinfo {volume} {7}},\ \bibinfo
  {pages} {1963--2001} (\bibinfo {year} {2006})}\BibitemShut {NoStop}%
\bibitem [{\citenamefont {Blachut}\ and\ \citenamefont
  {Gonz\'alez-Tokman}(2020)}]{Blachut2020}%
  \BibitemOpen
  \bibfield  {author} {\bibinfo {author} {\bibfnamefont {C.}~\bibnamefont
  {Blachut}}\ and\ \bibinfo {author} {\bibfnamefont {C.}~\bibnamefont
  {Gonz\'alez-Tokman}},\ }\bibfield  {title} {\enquote {\bibinfo {title} {A
  tale of two vortices: How numerical ergodic theory and transfer operators
  reveal fundamental changes to coherent structures in non-autonomous dynamical
  systems},}\ }\href {\doibase 10.3934/jcd.2020015} {\bibfield  {journal}
  {\bibinfo  {journal} {J. Comput. Dyn.}\ }\textbf {\bibinfo {volume} {7}},\
  \bibinfo {pages} {369--399} (\bibinfo {year} {2020})}\BibitemShut {NoStop}%
\bibitem [{\citenamefont {Ndour}, \citenamefont {Padberg-Gehle},\ and\
  \citenamefont {Rasmussen}(2021)}]{Ndour2021}%
  \BibitemOpen
  \bibfield  {author} {\bibinfo {author} {\bibfnamefont {M.}~\bibnamefont
  {Ndour}}, \bibinfo {author} {\bibfnamefont {K.}~\bibnamefont
  {Padberg-Gehle}}, \ and\ \bibinfo {author} {\bibfnamefont {M.}~\bibnamefont
  {Rasmussen}},\ }\bibfield  {title} {\enquote {\bibinfo {title} {Spectral
  early-warning signals for sudden changes in time-dependent flow patterns},}\
  }\href {\doibase 10.3390/fluids6020049} {\bibfield  {journal} {\bibinfo
  {journal} {Fluids}\ }\textbf {\bibinfo {volume} {6}},\ \bibinfo {pages} {49}
  (\bibinfo {year} {2021})}\BibitemShut {NoStop}%
\bibitem [{\citenamefont {Donner}\ \emph {et~al.}(2010)\citenamefont {Donner},
  \citenamefont {Zou}, \citenamefont {Donges}, \citenamefont {Marwan},\ and\
  \citenamefont {Kurths}}]{Donner2010}%
  \BibitemOpen
  \bibfield  {author} {\bibinfo {author} {\bibfnamefont {R.~V.}\ \bibnamefont
  {Donner}}, \bibinfo {author} {\bibfnamefont {Y.}~\bibnamefont {Zou}},
  \bibinfo {author} {\bibfnamefont {J.~F.}\ \bibnamefont {Donges}}, \bibinfo
  {author} {\bibfnamefont {N.}~\bibnamefont {Marwan}}, \ and\ \bibinfo {author}
  {\bibfnamefont {J.}~\bibnamefont {Kurths}},\ }\bibfield  {title} {\enquote
  {\bibinfo {title} {Ambiguities in recurrence-based complex network
  representations of time series},}\ }\href@noop {} {\bibfield  {journal}
  {\bibinfo  {journal} {Phys. Rev. E}\ }\textbf {\bibinfo {volume} {81}},\
  \bibinfo {pages} {015101} (\bibinfo {year} {2010})}\BibitemShut {NoStop}%
\end{thebibliography}%

\end{document}